\documentclass[12pt,leqno,fleqn]{article}
\usepackage{amsmath}
\usepackage{amsthm}
\usepackage{amssymb}
\usepackage{graphicx}
\usepackage{bbm}
\usepackage{float}
\usepackage{amsxtra}     
\usepackage{epsfig}
\usepackage{verbatim}
\usepackage[comma]{natbib}
\usepackage{multirow}
\usepackage{setspace}
\usepackage{comment}
\usepackage{graphics}
\usepackage{url}
\usepackage{enumerate}
\usepackage{setspace}
\usepackage{mathrsfs}
\usepackage{multirow}
\usepackage{booktabs}
\newcommand{\cprob}{\stackrel{\mathrm{P}}{\longrightarrow}}

\usepackage{amssymb}
\newcommand\Tau{\mathcal{T}}
\usepackage{xcolor,colortbl}
  \definecolor{unigreen}{rgb}{.1,.5,.25}
\DeclareMathOperator*{\argmin}{arg\,min}

\graphicspath{{figures/}}
\bibpunct{(}{)}{;}{a}{}{,}
\usepackage{color}
  \definecolor{snow}{rgb}{1.000000,0.980392,0.980392}
  \definecolor{ghost white}{rgb}{0.972549,0.972549,1.000000}
  \definecolor{GhostWhite}{rgb}{0.972549,0.972549,1.000000}
  \definecolor{white smoke}{rgb}{0.960784,0.960784,0.960784}
  \definecolor{WhiteSmoke}{rgb}{0.960784,0.960784,0.960784}
  \definecolor{gainsboro}{rgb}{0.862745,0.862745,0.862745}
  \definecolor{floral white}{rgb}{1.000000,0.980392,0.941176}
  \definecolor{FloralWhite}{rgb}{1.000000,0.980392,0.941176}
  \definecolor{old lace}{rgb}{0.992157,0.960784,0.901961}
  \definecolor{OldLace}{rgb}{0.992157,0.960784,0.901961}
  \definecolor{linen}{rgb}{0.980392,0.941176,0.901961}
  \definecolor{antique white}{rgb}{0.980392,0.921569,0.843137}
  \definecolor{AntiqueWhite}{rgb}{0.980392,0.921569,0.843137}
  \definecolor{papaya whip}{rgb}{1.000000,0.937255,0.835294}
  \definecolor{PapayaWhip}{rgb}{1.000000,0.937255,0.835294}
  \definecolor{blanched almond}{rgb}{1.000000,0.921569,0.803922}
  \definecolor{BlanchedAlmond}{rgb}{1.000000,0.921569,0.803922}
  \definecolor{bisque}{rgb}{1.000000,0.894118,0.768627}
  \definecolor{peach puff}{rgb}{1.000000,0.854902,0.725490}
  \definecolor{PeachPuff}{rgb}{1.000000,0.854902,0.725490}
  \definecolor{navajo white}{rgb}{1.000000,0.870588,0.678431}
  \definecolor{NavajoWhite}{rgb}{1.000000,0.870588,0.678431}
  \definecolor{moccasin}{rgb}{1.000000,0.894118,0.709804}
  \definecolor{cornsilk}{rgb}{1.000000,0.972549,0.862745}
  \definecolor{ivory}{rgb}{1.000000,1.000000,0.941176}
  \definecolor{lemon chiffon}{rgb}{1.000000,0.980392,0.803922}
  \definecolor{LemonChiffon}{rgb}{1.000000,0.980392,0.803922}
  \definecolor{seashell}{rgb}{1.000000,0.960784,0.933333}
  \definecolor{honeydew}{rgb}{0.941176,1.000000,0.941176}
  \definecolor{mint cream}{rgb}{0.960784,1.000000,0.980392}
  \definecolor{MintCream}{rgb}{0.960784,1.000000,0.980392}
  \definecolor{azure}{rgb}{0.941176,1.000000,1.000000}
  \definecolor{alice blue}{rgb}{0.941176,0.972549,1.000000}
  \definecolor{AliceBlue}{rgb}{0.941176,0.972549,1.000000}
  \definecolor{lavender}{rgb}{0.901961,0.901961,0.980392}
  \definecolor{lavender blush}{rgb}{1.000000,0.941176,0.960784}
  \definecolor{LavenderBlush}{rgb}{1.000000,0.941176,0.960784}
  \definecolor{misty rose}{rgb}{1.000000,0.894118,0.882353}
  \definecolor{MistyRose}{rgb}{1.000000,0.894118,0.882353}
  \definecolor{white}{rgb}{1.000000,1.000000,1.000000}
  \definecolor{black}{rgb}{0.000000,0.000000,0.000000}
  \definecolor{dark slate gray}{rgb}{0.184314,0.309804,0.309804}
  \definecolor{DarkSlateGray}{rgb}{0.184314,0.309804,0.309804}
  \definecolor{dark slate grey}{rgb}{0.184314,0.309804,0.309804}
  \definecolor{DarkSlateGrey}{rgb}{0.184314,0.309804,0.309804}
  \definecolor{dim gray}{rgb}{0.411765,0.411765,0.411765}
  \definecolor{DimGray}{rgb}{0.411765,0.411765,0.411765}
  \definecolor{dim grey}{rgb}{0.411765,0.411765,0.411765}
  \definecolor{DimGrey}{rgb}{0.411765,0.411765,0.411765}
  \definecolor{slate gray}{rgb}{0.439216,0.501961,0.564706}
  \definecolor{SlateGray}{rgb}{0.439216,0.501961,0.564706}
  \definecolor{slate grey}{rgb}{0.439216,0.501961,0.564706}
  \definecolor{SlateGrey}{rgb}{0.439216,0.501961,0.564706}
  \definecolor{light slate gray}{rgb}{0.466667,0.533333,0.600000}
  \definecolor{LightSlateGray}{rgb}{0.466667,0.533333,0.600000}
  \definecolor{light slate grey}{rgb}{0.466667,0.533333,0.600000}
  \definecolor{LightSlateGrey}{rgb}{0.466667,0.533333,0.600000}
  \definecolor{gray}{rgb}{0.745098,0.745098,0.745098}
  \definecolor{grey}{rgb}{0.745098,0.745098,0.745098}
  \definecolor{light grey}{rgb}{0.827451,0.827451,0.827451}
  \definecolor{LightGrey}{rgb}{0.827451,0.827451,0.827451}
  \definecolor{light gray}{rgb}{0.827451,0.827451,0.827451}
  \definecolor{LightGray}{rgb}{0.827451,0.827451,0.827451}
  \definecolor{midnight blue}{rgb}{0.098039,0.098039,0.439216}
  \definecolor{MidnightBlue}{rgb}{0.098039,0.098039,0.439216}
  \definecolor{navy}{rgb}{0.000000,0.000000,0.501961}
  \definecolor{navy blue}{rgb}{0.000000,0.000000,0.501961}
  \definecolor{NavyBlue}{rgb}{0.000000,0.000000,0.501961}
  \definecolor{cornflower blue}{rgb}{0.392157,0.584314,0.929412}
  \definecolor{CornflowerBlue}{rgb}{0.392157,0.584314,0.929412}
  \definecolor{dark slate blue}{rgb}{0.282353,0.239216,0.545098}
  \definecolor{DarkSlateBlue}{rgb}{0.282353,0.239216,0.545098}
  \definecolor{slate blue}{rgb}{0.415686,0.352941,0.803922}
  \definecolor{SlateBlue}{rgb}{0.415686,0.352941,0.803922}
  \definecolor{medium slate blue}{rgb}{0.482353,0.407843,0.933333}
  \definecolor{MediumSlateBlue}{rgb}{0.482353,0.407843,0.933333}
  \definecolor{light slate blue}{rgb}{0.517647,0.439216,1.000000}
  \definecolor{LightSlateBlue}{rgb}{0.517647,0.439216,1.000000}
  \definecolor{medium blue}{rgb}{0.000000,0.000000,0.803922}
  \definecolor{MediumBlue}{rgb}{0.000000,0.000000,0.803922}
  \definecolor{royal blue}{rgb}{0.254902,0.411765,0.882353}
  \definecolor{RoyalBlue}{rgb}{0.254902,0.411765,0.882353}
  \definecolor{blue}{rgb}{0.000000,0.000000,1.000000}
  \definecolor{dodger blue}{rgb}{0.117647,0.564706,1.000000}
  \definecolor{DodgerBlue}{rgb}{0.117647,0.564706,1.000000}
  \definecolor{deep sky blue}{rgb}{0.000000,0.749020,1.000000}
  \definecolor{DeepSkyBlue}{rgb}{0.000000,0.749020,1.000000}
  \definecolor{sky blue}{rgb}{0.529412,0.807843,0.921569}
  \definecolor{SkyBlue}{rgb}{0.529412,0.807843,0.921569}
  \definecolor{light sky blue}{rgb}{0.529412,0.807843,0.980392}
  \definecolor{LightSkyBlue}{rgb}{0.529412,0.807843,0.980392}
  \definecolor{steel blue}{rgb}{0.274510,0.509804,0.705882}
  \definecolor{SteelBlue}{rgb}{0.274510,0.509804,0.705882}
  \definecolor{light steel blue}{rgb}{0.690196,0.768627,0.870588}
  \definecolor{LightSteelBlue}{rgb}{0.690196,0.768627,0.870588}
  \definecolor{light blue}{rgb}{0.678431,0.847059,0.901961}
  \definecolor{LightBlue}{rgb}{0.678431,0.847059,0.901961}
  \definecolor{powder blue}{rgb}{0.690196,0.878431,0.901961}
  \definecolor{PowderBlue}{rgb}{0.690196,0.878431,0.901961}
  \definecolor{pale turquoise}{rgb}{0.686275,0.933333,0.933333}
  \definecolor{PaleTurquoise}{rgb}{0.686275,0.933333,0.933333}
  \definecolor{dark turquoise}{rgb}{0.000000,0.807843,0.819608}
  \definecolor{DarkTurquoise}{rgb}{0.000000,0.807843,0.819608}
  \definecolor{medium turquoise}{rgb}{0.282353,0.819608,0.800000}
  \definecolor{MediumTurquoise}{rgb}{0.282353,0.819608,0.800000}
  \definecolor{turquoise}{rgb}{0.250980,0.878431,0.815686}
  \definecolor{cyan}{rgb}{0.000000,1.000000,1.000000}
  \definecolor{light cyan}{rgb}{0.878431,1.000000,1.000000}
  \definecolor{LightCyan}{rgb}{0.878431,1.000000,1.000000}
  \definecolor{cadet blue}{rgb}{0.372549,0.619608,0.627451}
  \definecolor{CadetBlue}{rgb}{0.372549,0.619608,0.627451}
  \definecolor{medium aquamarine}{rgb}{0.400000,0.803922,0.666667}
  \definecolor{MediumAquamarine}{rgb}{0.400000,0.803922,0.666667}
  \definecolor{aquamarine}{rgb}{0.498039,1.000000,0.831373}
  \definecolor{dark green}{rgb}{0.000000,0.392157,0.000000}
  \definecolor{DarkGreen}{rgb}{0.000000,0.392157,0.000000}
  \definecolor{dark olive green}{rgb}{0.333333,0.419608,0.184314}
  \definecolor{DarkOliveGreen}{rgb}{0.333333,0.419608,0.184314}
  \definecolor{dark sea green}{rgb}{0.560784,0.737255,0.560784}
  \definecolor{DarkSeaGreen}{rgb}{0.560784,0.737255,0.560784}
  \definecolor{sea green}{rgb}{0.180392,0.545098,0.341176}
  \definecolor{SeaGreen}{rgb}{0.180392,0.545098,0.341176}
  \definecolor{medium sea green}{rgb}{0.235294,0.701961,0.443137}
  \definecolor{MediumSeaGreen}{rgb}{0.235294,0.701961,0.443137}
  \definecolor{light sea green}{rgb}{0.125490,0.698039,0.666667}
  \definecolor{LightSeaGreen}{rgb}{0.125490,0.698039,0.666667}
  \definecolor{pale green}{rgb}{0.596078,0.984314,0.596078}
  \definecolor{PaleGreen}{rgb}{0.596078,0.984314,0.596078}
  \definecolor{spring green}{rgb}{0.000000,1.000000,0.498039}
  \definecolor{SpringGreen}{rgb}{0.000000,1.000000,0.498039}
  \definecolor{lawn green}{rgb}{0.486275,0.988235,0.000000}
  \definecolor{LawnGreen}{rgb}{0.486275,0.988235,0.000000}
  \definecolor{green}{rgb}{0.000000,1.000000,0.000000}
  \definecolor{chartreuse}{rgb}{0.498039,1.000000,0.000000}
  \definecolor{medium spring green}{rgb}{0.000000,0.980392,0.603922}
  \definecolor{MediumSpringGreen}{rgb}{0.000000,0.980392,0.603922}
  \definecolor{green yellow}{rgb}{0.678431,1.000000,0.184314}
  \definecolor{GreenYellow}{rgb}{0.678431,1.000000,0.184314}
  \definecolor{lime green}{rgb}{0.196078,0.803922,0.196078}
  \definecolor{LimeGreen}{rgb}{0.196078,0.803922,0.196078}
  \definecolor{yellow green}{rgb}{0.603922,0.803922,0.196078}
  \definecolor{YellowGreen}{rgb}{0.603922,0.803922,0.196078}
  \definecolor{forest green}{rgb}{0.133333,0.545098,0.133333}
  \definecolor{ForestGreen}{rgb}{0.133333,0.545098,0.133333}
  \definecolor{olive drab}{rgb}{0.419608,0.556863,0.137255}
  \definecolor{OliveDrab}{rgb}{0.419608,0.556863,0.137255}
  \definecolor{dark khaki}{rgb}{0.741176,0.717647,0.419608}
  \definecolor{DarkKhaki}{rgb}{0.741176,0.717647,0.419608}
  \definecolor{khaki}{rgb}{0.941176,0.901961,0.549020}
  \definecolor{pale goldenrod}{rgb}{0.933333,0.909804,0.666667}
  \definecolor{PaleGoldenrod}{rgb}{0.933333,0.909804,0.666667}
  \definecolor{light goldenrod yellow}{rgb}{0.980392,0.980392,0.823529}
  \definecolor{LightGoldenrodYellow}{rgb}{0.980392,0.980392,0.823529}
  \definecolor{light yellow}{rgb}{1.000000,1.000000,0.878431}
  \definecolor{LightYellow}{rgb}{1.000000,1.000000,0.878431}
  \definecolor{yellow}{rgb}{1.000000,1.000000,0.000000}
  \definecolor{gold}{rgb}{1.000000,0.843137,0.000000}
  \definecolor{light goldenrod}{rgb}{0.933333,0.866667,0.509804}
  \definecolor{LightGoldenrod}{rgb}{0.933333,0.866667,0.509804}
  \definecolor{goldenrod}{rgb}{0.854902,0.647059,0.125490}
  \definecolor{dark goldenrod}{rgb}{0.721569,0.525490,0.043137}
  \definecolor{DarkGoldenrod}{rgb}{0.721569,0.525490,0.043137}
  \definecolor{rosy brown}{rgb}{0.737255,0.560784,0.560784}
  \definecolor{RosyBrown}{rgb}{0.737255,0.560784,0.560784}
  \definecolor{indian red}{rgb}{0.803922,0.360784,0.360784}
  \definecolor{IndianRed}{rgb}{0.803922,0.360784,0.360784}
  \definecolor{saddle brown}{rgb}{0.545098,0.270588,0.074510}
  \definecolor{SaddleBrown}{rgb}{0.545098,0.270588,0.074510}
  \definecolor{sienna}{rgb}{0.627451,0.321569,0.176471}
  \definecolor{peru}{rgb}{0.803922,0.521569,0.247059}
  \definecolor{burlywood}{rgb}{0.870588,0.721569,0.529412}
  \definecolor{beige}{rgb}{0.960784,0.960784,0.862745}
  \definecolor{wheat}{rgb}{0.960784,0.870588,0.701961}
  \definecolor{sandy brown}{rgb}{0.956863,0.643137,0.376471}
  \definecolor{SandyBrown}{rgb}{0.956863,0.643137,0.376471}
  \definecolor{tan}{rgb}{0.823529,0.705882,0.549020}
  \definecolor{chocolate}{rgb}{0.823529,0.411765,0.117647}
  \definecolor{firebrick}{rgb}{0.698039,0.133333,0.133333}
  \definecolor{brown}{rgb}{0.647059,0.164706,0.164706}
  \definecolor{dark salmon}{rgb}{0.913725,0.588235,0.478431}
  \definecolor{DarkSalmon}{rgb}{0.913725,0.588235,0.478431}
  \definecolor{salmon}{rgb}{0.980392,0.501961,0.447059}
  \definecolor{light salmon}{rgb}{1.000000,0.627451,0.478431}
  \definecolor{LightSalmon}{rgb}{1.000000,0.627451,0.478431}
  \definecolor{orange}{rgb}{1.000000,0.647059,0.000000}
  \definecolor{dark orange}{rgb}{1.000000,0.549020,0.000000}
  \definecolor{DarkOrange}{rgb}{1.000000,0.549020,0.000000}
  \definecolor{coral}{rgb}{1.000000,0.498039,0.313726}
  \definecolor{light coral}{rgb}{0.941176,0.501961,0.501961}
  \definecolor{LightCoral}{rgb}{0.941176,0.501961,0.501961}
  \definecolor{tomato}{rgb}{1.000000,0.388235,0.278431}
  \definecolor{orange red}{rgb}{1.000000,0.270588,0.000000}
  \definecolor{OrangeRed}{rgb}{1.000000,0.270588,0.000000}
  \definecolor{red}{rgb}{1.000000,0.000000,0.000000}
  \definecolor{hot pink}{rgb}{1.000000,0.411765,0.705882}
  \definecolor{HotPink}{rgb}{1.000000,0.411765,0.705882}
  \definecolor{deep pink}{rgb}{1.000000,0.078431,0.576471}
  \definecolor{DeepPink}{rgb}{1.000000,0.078431,0.576471}
  \definecolor{pink}{rgb}{1.000000,0.752941,0.796078}
  \definecolor{light pink}{rgb}{1.000000,0.713726,0.756863}
  \definecolor{LightPink}{rgb}{1.000000,0.713726,0.756863}
  \definecolor{pale violet red}{rgb}{0.858824,0.439216,0.576471}
  \definecolor{PaleVioletRed}{rgb}{0.858824,0.439216,0.576471}
  \definecolor{maroon}{rgb}{0.690196,0.188235,0.376471}
  \definecolor{medium violet red}{rgb}{0.780392,0.082353,0.521569}
  \definecolor{MediumVioletRed}{rgb}{0.780392,0.082353,0.521569}
  \definecolor{violet red}{rgb}{0.815686,0.125490,0.564706}
  \definecolor{VioletRed}{rgb}{0.815686,0.125490,0.564706}
  \definecolor{magenta}{rgb}{1.000000,0.000000,1.000000}
  \definecolor{violet}{rgb}{0.933333,0.509804,0.933333}
  \definecolor{plum}{rgb}{0.866667,0.627451,0.866667}
  \definecolor{orchid}{rgb}{0.854902,0.439216,0.839216}
  \definecolor{medium orchid}{rgb}{0.729412,0.333333,0.827451}
  \definecolor{MediumOrchid}{rgb}{0.729412,0.333333,0.827451}
  \definecolor{dark orchid}{rgb}{0.600000,0.196078,0.800000}
  \definecolor{DarkOrchid}{rgb}{0.600000,0.196078,0.800000}
  \definecolor{dark violet}{rgb}{0.580392,0.000000,0.827451}
  \definecolor{DarkViolet}{rgb}{0.580392,0.000000,0.827451}
  \definecolor{blue violet}{rgb}{0.541176,0.168627,0.886275}
  \definecolor{BlueViolet}{rgb}{0.541176,0.168627,0.886275}
  \definecolor{purple}{rgb}{0.627451,0.125490,0.941176}
  \definecolor{medium purple}{rgb}{0.576471,0.439216,0.858824}
  \definecolor{MediumPurple}{rgb}{0.576471,0.439216,0.858824}
  \definecolor{thistle}{rgb}{0.847059,0.749020,0.847059}
  \definecolor{snow1}{rgb}{1.000000,0.980392,0.980392}
  \definecolor{snow2}{rgb}{0.933333,0.913725,0.913725}
  \definecolor{snow3}{rgb}{0.803922,0.788235,0.788235}
  \definecolor{snow4}{rgb}{0.545098,0.537255,0.537255}
  \definecolor{seashell1}{rgb}{1.000000,0.960784,0.933333}
  \definecolor{seashell2}{rgb}{0.933333,0.898039,0.870588}
  \definecolor{seashell3}{rgb}{0.803922,0.772549,0.749020}
  \definecolor{seashell4}{rgb}{0.545098,0.525490,0.509804}
  \definecolor{AntiqueWhite1}{rgb}{1.000000,0.937255,0.858824}
  \definecolor{AntiqueWhite2}{rgb}{0.933333,0.874510,0.800000}
  \definecolor{AntiqueWhite3}{rgb}{0.803922,0.752941,0.690196}
  \definecolor{AntiqueWhite4}{rgb}{0.545098,0.513726,0.470588}
  \definecolor{bisque1}{rgb}{1.000000,0.894118,0.768627}
  \definecolor{bisque2}{rgb}{0.933333,0.835294,0.717647}
  \definecolor{bisque3}{rgb}{0.803922,0.717647,0.619608}
  \definecolor{bisque4}{rgb}{0.545098,0.490196,0.419608}
  \definecolor{PeachPuff1}{rgb}{1.000000,0.854902,0.725490}
  \definecolor{PeachPuff2}{rgb}{0.933333,0.796078,0.678431}
  \definecolor{PeachPuff3}{rgb}{0.803922,0.686275,0.584314}
  \definecolor{PeachPuff4}{rgb}{0.545098,0.466667,0.396078}
  \definecolor{NavajoWhite1}{rgb}{1.000000,0.870588,0.678431}
  \definecolor{NavajoWhite2}{rgb}{0.933333,0.811765,0.631373}
  \definecolor{NavajoWhite3}{rgb}{0.803922,0.701961,0.545098}
  \definecolor{NavajoWhite4}{rgb}{0.545098,0.474510,0.368627}
  \definecolor{LemonChiffon1}{rgb}{1.000000,0.980392,0.803922}
  \definecolor{LemonChiffon2}{rgb}{0.933333,0.913725,0.749020}
  \definecolor{LemonChiffon3}{rgb}{0.803922,0.788235,0.647059}
  \definecolor{LemonChiffon4}{rgb}{0.545098,0.537255,0.439216}
  \definecolor{cornsilk1}{rgb}{1.000000,0.972549,0.862745}
  \definecolor{cornsilk2}{rgb}{0.933333,0.909804,0.803922}
  \definecolor{cornsilk3}{rgb}{0.803922,0.784314,0.694118}
  \definecolor{cornsilk4}{rgb}{0.545098,0.533333,0.470588}
  \definecolor{ivory1}{rgb}{1.000000,1.000000,0.941176}
  \definecolor{ivory2}{rgb}{0.933333,0.933333,0.878431}
  \definecolor{ivory3}{rgb}{0.803922,0.803922,0.756863}
  \definecolor{ivory4}{rgb}{0.545098,0.545098,0.513726}
  \definecolor{honeydew1}{rgb}{0.941176,1.000000,0.941176}
  \definecolor{honeydew2}{rgb}{0.878431,0.933333,0.878431}
  \definecolor{honeydew3}{rgb}{0.756863,0.803922,0.756863}
  \definecolor{honeydew4}{rgb}{0.513726,0.545098,0.513726}
  \definecolor{LavenderBlush1}{rgb}{1.000000,0.941176,0.960784}
  \definecolor{LavenderBlush2}{rgb}{0.933333,0.878431,0.898039}
  \definecolor{LavenderBlush3}{rgb}{0.803922,0.756863,0.772549}
  \definecolor{LavenderBlush4}{rgb}{0.545098,0.513726,0.525490}
  \definecolor{MistyRose1}{rgb}{1.000000,0.894118,0.882353}
  \definecolor{MistyRose2}{rgb}{0.933333,0.835294,0.823529}
  \definecolor{MistyRose3}{rgb}{0.803922,0.717647,0.709804}
  \definecolor{MistyRose4}{rgb}{0.545098,0.490196,0.482353}
  \definecolor{azure1}{rgb}{0.941176,1.000000,1.000000}
  \definecolor{azure2}{rgb}{0.878431,0.933333,0.933333}
  \definecolor{azure3}{rgb}{0.756863,0.803922,0.803922}
  \definecolor{azure4}{rgb}{0.513726,0.545098,0.545098}
  \definecolor{SlateBlue1}{rgb}{0.513726,0.435294,1.000000}
  \definecolor{SlateBlue2}{rgb}{0.478431,0.403922,0.933333}
  \definecolor{SlateBlue3}{rgb}{0.411765,0.349020,0.803922}
  \definecolor{SlateBlue4}{rgb}{0.278431,0.235294,0.545098}
  \definecolor{RoyalBlue1}{rgb}{0.282353,0.462745,1.000000}
  \definecolor{RoyalBlue2}{rgb}{0.262745,0.431373,0.933333}
  \definecolor{RoyalBlue3}{rgb}{0.227451,0.372549,0.803922}
  \definecolor{RoyalBlue4}{rgb}{0.152941,0.250980,0.545098}
  \definecolor{blue1}{rgb}{0.000000,0.000000,1.000000}
  \definecolor{blue2}{rgb}{0.000000,0.000000,0.933333}
  \definecolor{blue3}{rgb}{0.000000,0.000000,0.803922}
  \definecolor{blue4}{rgb}{0.000000,0.000000,0.545098}
  \definecolor{DodgerBlue1}{rgb}{0.117647,0.564706,1.000000}
  \definecolor{DodgerBlue2}{rgb}{0.109804,0.525490,0.933333}
  \definecolor{DodgerBlue3}{rgb}{0.094118,0.454902,0.803922}
  \definecolor{DodgerBlue4}{rgb}{0.062745,0.305882,0.545098}
  \definecolor{SteelBlue1}{rgb}{0.388235,0.721569,1.000000}
  \definecolor{SteelBlue2}{rgb}{0.360784,0.674510,0.933333}
  \definecolor{SteelBlue3}{rgb}{0.309804,0.580392,0.803922}
  \definecolor{SteelBlue4}{rgb}{0.211765,0.392157,0.545098}
  \definecolor{DeepSkyBlue1}{rgb}{0.000000,0.749020,1.000000}
  \definecolor{DeepSkyBlue2}{rgb}{0.000000,0.698039,0.933333}
  \definecolor{DeepSkyBlue3}{rgb}{0.000000,0.603922,0.803922}
  \definecolor{DeepSkyBlue4}{rgb}{0.000000,0.407843,0.545098}
  \definecolor{SkyBlue1}{rgb}{0.529412,0.807843,1.000000}
  \definecolor{SkyBlue2}{rgb}{0.494118,0.752941,0.933333}
  \definecolor{SkyBlue3}{rgb}{0.423529,0.650980,0.803922}
  \definecolor{SkyBlue4}{rgb}{0.290196,0.439216,0.545098}
  \definecolor{LightSkyBlue1}{rgb}{0.690196,0.886275,1.000000}
  \definecolor{LightSkyBlue2}{rgb}{0.643137,0.827451,0.933333}
  \definecolor{LightSkyBlue3}{rgb}{0.552941,0.713726,0.803922}
  \definecolor{LightSkyBlue4}{rgb}{0.376471,0.482353,0.545098}
  \definecolor{SlateGray1}{rgb}{0.776471,0.886275,1.000000}
  \definecolor{SlateGray2}{rgb}{0.725490,0.827451,0.933333}
  \definecolor{SlateGray3}{rgb}{0.623529,0.713726,0.803922}
  \definecolor{SlateGray4}{rgb}{0.423529,0.482353,0.545098}
  \definecolor{LightSteelBlue1}{rgb}{0.792157,0.882353,1.000000}
  \definecolor{LightSteelBlue2}{rgb}{0.737255,0.823529,0.933333}
  \definecolor{LightSteelBlue3}{rgb}{0.635294,0.709804,0.803922}
  \definecolor{LightSteelBlue4}{rgb}{0.431373,0.482353,0.545098}
  \definecolor{LightBlue1}{rgb}{0.749020,0.937255,1.000000}
  \definecolor{LightBlue2}{rgb}{0.698039,0.874510,0.933333}
  \definecolor{LightBlue3}{rgb}{0.603922,0.752941,0.803922}
  \definecolor{LightBlue4}{rgb}{0.407843,0.513726,0.545098}
  \definecolor{LightCyan1}{rgb}{0.878431,1.000000,1.000000}
  \definecolor{LightCyan2}{rgb}{0.819608,0.933333,0.933333}
  \definecolor{LightCyan3}{rgb}{0.705882,0.803922,0.803922}
  \definecolor{LightCyan4}{rgb}{0.478431,0.545098,0.545098}
  \definecolor{PaleTurquoise1}{rgb}{0.733333,1.000000,1.000000}
  \definecolor{PaleTurquoise2}{rgb}{0.682353,0.933333,0.933333}
  \definecolor{PaleTurquoise3}{rgb}{0.588235,0.803922,0.803922}
  \definecolor{PaleTurquoise4}{rgb}{0.400000,0.545098,0.545098}
  \definecolor{CadetBlue1}{rgb}{0.596078,0.960784,1.000000}
  \definecolor{CadetBlue2}{rgb}{0.556863,0.898039,0.933333}
  \definecolor{CadetBlue3}{rgb}{0.478431,0.772549,0.803922}
  \definecolor{CadetBlue4}{rgb}{0.325490,0.525490,0.545098}
  \definecolor{turquoise1}{rgb}{0.000000,0.960784,1.000000}
  \definecolor{turquoise2}{rgb}{0.000000,0.898039,0.933333}
  \definecolor{turquoise3}{rgb}{0.000000,0.772549,0.803922}
  \definecolor{turquoise4}{rgb}{0.000000,0.525490,0.545098}
  \definecolor{cyan1}{rgb}{0.000000,1.000000,1.000000}
  \definecolor{cyan2}{rgb}{0.000000,0.933333,0.933333}
  \definecolor{cyan3}{rgb}{0.000000,0.803922,0.803922}
  \definecolor{cyan4}{rgb}{0.000000,0.545098,0.545098}
  \definecolor{DarkSlateGray1}{rgb}{0.592157,1.000000,1.000000}
  \definecolor{DarkSlateGray2}{rgb}{0.552941,0.933333,0.933333}
  \definecolor{DarkSlateGray3}{rgb}{0.474510,0.803922,0.803922}
  \definecolor{DarkSlateGray4}{rgb}{0.321569,0.545098,0.545098}
  \definecolor{aquamarine1}{rgb}{0.498039,1.000000,0.831373}
  \definecolor{aquamarine2}{rgb}{0.462745,0.933333,0.776471}
  \definecolor{aquamarine3}{rgb}{0.400000,0.803922,0.666667}
  \definecolor{aquamarine4}{rgb}{0.270588,0.545098,0.454902}
  \definecolor{DarkSeaGreen1}{rgb}{0.756863,1.000000,0.756863}
  \definecolor{DarkSeaGreen2}{rgb}{0.705882,0.933333,0.705882}
  \definecolor{DarkSeaGreen3}{rgb}{0.607843,0.803922,0.607843}
  \definecolor{DarkSeaGreen4}{rgb}{0.411765,0.545098,0.411765}
  \definecolor{SeaGreen1}{rgb}{0.329412,1.000000,0.623529}
  \definecolor{SeaGreen2}{rgb}{0.305882,0.933333,0.580392}
  \definecolor{SeaGreen3}{rgb}{0.262745,0.803922,0.501961}
  \definecolor{SeaGreen4}{rgb}{0.180392,0.545098,0.341176}
  \definecolor{PaleGreen1}{rgb}{0.603922,1.000000,0.603922}
  \definecolor{PaleGreen2}{rgb}{0.564706,0.933333,0.564706}
  \definecolor{PaleGreen3}{rgb}{0.486275,0.803922,0.486275}
  \definecolor{PaleGreen4}{rgb}{0.329412,0.545098,0.329412}
  \definecolor{SpringGreen1}{rgb}{0.000000,1.000000,0.498039}
  \definecolor{SpringGreen2}{rgb}{0.000000,0.933333,0.462745}
  \definecolor{SpringGreen3}{rgb}{0.000000,0.803922,0.400000}
  \definecolor{SpringGreen4}{rgb}{0.000000,0.545098,0.270588}
  \definecolor{green1}{rgb}{0.000000,1.000000,0.000000}
  \definecolor{green2}{rgb}{0.000000,0.933333,0.000000}
  \definecolor{green3}{rgb}{0.000000,0.803922,0.000000}
  \definecolor{green4}{rgb}{0.000000,0.545098,0.000000}
  \definecolor{chartreuse1}{rgb}{0.498039,1.000000,0.000000}
  \definecolor{chartreuse2}{rgb}{0.462745,0.933333,0.000000}
  \definecolor{chartreuse3}{rgb}{0.400000,0.803922,0.000000}
  \definecolor{chartreuse4}{rgb}{0.270588,0.545098,0.000000}
  \definecolor{OliveDrab1}{rgb}{0.752941,1.000000,0.243137}
  \definecolor{OliveDrab2}{rgb}{0.701961,0.933333,0.227451}
  \definecolor{OliveDrab3}{rgb}{0.603922,0.803922,0.196078}
  \definecolor{OliveDrab4}{rgb}{0.411765,0.545098,0.133333}
  \definecolor{DarkOliveGreen1}{rgb}{0.792157,1.000000,0.439216}
  \definecolor{DarkOliveGreen2}{rgb}{0.737255,0.933333,0.407843}
  \definecolor{DarkOliveGreen3}{rgb}{0.635294,0.803922,0.352941}
  \definecolor{DarkOliveGreen4}{rgb}{0.431373,0.545098,0.239216}
  \definecolor{khaki1}{rgb}{1.000000,0.964706,0.560784}
  \definecolor{khaki2}{rgb}{0.933333,0.901961,0.521569}
  \definecolor{khaki3}{rgb}{0.803922,0.776471,0.450980}
  \definecolor{khaki4}{rgb}{0.545098,0.525490,0.305882}
  \definecolor{LightGoldenrod1}{rgb}{1.000000,0.925490,0.545098}
  \definecolor{LightGoldenrod2}{rgb}{0.933333,0.862745,0.509804}
  \definecolor{LightGoldenrod3}{rgb}{0.803922,0.745098,0.439216}
  \definecolor{LightGoldenrod4}{rgb}{0.545098,0.505882,0.298039}
  \definecolor{LightYellow1}{rgb}{1.000000,1.000000,0.878431}
  \definecolor{LightYellow2}{rgb}{0.933333,0.933333,0.819608}
  \definecolor{LightYellow3}{rgb}{0.803922,0.803922,0.705882}
  \definecolor{LightYellow4}{rgb}{0.545098,0.545098,0.478431}
  \definecolor{yellow1}{rgb}{1.000000,1.000000,0.000000}
  \definecolor{yellow2}{rgb}{0.933333,0.933333,0.000000}
  \definecolor{yellow3}{rgb}{0.803922,0.803922,0.000000}
  \definecolor{yellow4}{rgb}{0.545098,0.545098,0.000000}
  \definecolor{gold1}{rgb}{1.000000,0.843137,0.000000}
  \definecolor{gold2}{rgb}{0.933333,0.788235,0.000000}
  \definecolor{gold3}{rgb}{0.803922,0.678431,0.000000}
  \definecolor{gold4}{rgb}{0.545098,0.458824,0.000000}
  \definecolor{goldenrod1}{rgb}{1.000000,0.756863,0.145098}
  \definecolor{goldenrod2}{rgb}{0.933333,0.705882,0.133333}
  \definecolor{goldenrod3}{rgb}{0.803922,0.607843,0.113725}
  \definecolor{goldenrod4}{rgb}{0.545098,0.411765,0.078431}
  \definecolor{DarkGoldenrod1}{rgb}{1.000000,0.725490,0.058824}
  \definecolor{DarkGoldenrod2}{rgb}{0.933333,0.678431,0.054902}
  \definecolor{DarkGoldenrod3}{rgb}{0.803922,0.584314,0.047059}
  \definecolor{DarkGoldenrod4}{rgb}{0.545098,0.396078,0.031373}
  \definecolor{RosyBrown1}{rgb}{1.000000,0.756863,0.756863}
  \definecolor{RosyBrown2}{rgb}{0.933333,0.705882,0.705882}
  \definecolor{RosyBrown3}{rgb}{0.803922,0.607843,0.607843}
  \definecolor{RosyBrown4}{rgb}{0.545098,0.411765,0.411765}
  \definecolor{IndianRed1}{rgb}{1.000000,0.415686,0.415686}
  \definecolor{IndianRed2}{rgb}{0.933333,0.388235,0.388235}
  \definecolor{IndianRed3}{rgb}{0.803922,0.333333,0.333333}
  \definecolor{IndianRed4}{rgb}{0.545098,0.227451,0.227451}
  \definecolor{sienna1}{rgb}{1.000000,0.509804,0.278431}
  \definecolor{sienna2}{rgb}{0.933333,0.474510,0.258824}
  \definecolor{sienna3}{rgb}{0.803922,0.407843,0.223529}
  \definecolor{sienna4}{rgb}{0.545098,0.278431,0.149020}
  \definecolor{burlywood1}{rgb}{1.000000,0.827451,0.607843}
  \definecolor{burlywood2}{rgb}{0.933333,0.772549,0.568627}
  \definecolor{burlywood3}{rgb}{0.803922,0.666667,0.490196}
  \definecolor{burlywood4}{rgb}{0.545098,0.450980,0.333333}
  \definecolor{wheat1}{rgb}{1.000000,0.905882,0.729412}
  \definecolor{wheat2}{rgb}{0.933333,0.847059,0.682353}
  \definecolor{wheat3}{rgb}{0.803922,0.729412,0.588235}
  \definecolor{wheat4}{rgb}{0.545098,0.494118,0.400000}
  \definecolor{tan1}{rgb}{1.000000,0.647059,0.309804}
  \definecolor{tan2}{rgb}{0.933333,0.603922,0.286275}
  \definecolor{tan3}{rgb}{0.803922,0.521569,0.247059}
  \definecolor{tan4}{rgb}{0.545098,0.352941,0.168627}
  \definecolor{chocolate1}{rgb}{1.000000,0.498039,0.141176}
  \definecolor{chocolate2}{rgb}{0.933333,0.462745,0.129412}
  \definecolor{chocolate3}{rgb}{0.803922,0.400000,0.113725}
  \definecolor{chocolate4}{rgb}{0.545098,0.270588,0.074510}
  \definecolor{firebrick1}{rgb}{1.000000,0.188235,0.188235}
  \definecolor{firebrick2}{rgb}{0.933333,0.172549,0.172549}
  \definecolor{firebrick3}{rgb}{0.803922,0.149020,0.149020}
  \definecolor{firebrick4}{rgb}{0.545098,0.101961,0.101961}
  \definecolor{brown1}{rgb}{1.000000,0.250980,0.250980}
  \definecolor{brown2}{rgb}{0.933333,0.231373,0.231373}
  \definecolor{brown3}{rgb}{0.803922,0.200000,0.200000}
  \definecolor{brown4}{rgb}{0.545098,0.137255,0.137255}
  \definecolor{salmon1}{rgb}{1.000000,0.549020,0.411765}
  \definecolor{salmon2}{rgb}{0.933333,0.509804,0.384314}
  \definecolor{salmon3}{rgb}{0.803922,0.439216,0.329412}
  \definecolor{salmon4}{rgb}{0.545098,0.298039,0.223529}
  \definecolor{LightSalmon1}{rgb}{1.000000,0.627451,0.478431}
  \definecolor{LightSalmon2}{rgb}{0.933333,0.584314,0.447059}
  \definecolor{LightSalmon3}{rgb}{0.803922,0.505882,0.384314}
  \definecolor{LightSalmon4}{rgb}{0.545098,0.341176,0.258824}
  \definecolor{orange1}{rgb}{1.000000,0.647059,0.000000}
  \definecolor{orange2}{rgb}{0.933333,0.603922,0.000000}
  \definecolor{orange3}{rgb}{0.803922,0.521569,0.000000}
  \definecolor{orange4}{rgb}{0.545098,0.352941,0.000000}
  \definecolor{DarkOrange1}{rgb}{1.000000,0.498039,0.000000}
  \definecolor{DarkOrange2}{rgb}{0.933333,0.462745,0.000000}
  \definecolor{DarkOrange3}{rgb}{0.803922,0.400000,0.000000}
  \definecolor{DarkOrange4}{rgb}{0.545098,0.270588,0.000000}
  \definecolor{coral1}{rgb}{1.000000,0.447059,0.337255}
  \definecolor{coral2}{rgb}{0.933333,0.415686,0.313726}
  \definecolor{coral3}{rgb}{0.803922,0.356863,0.270588}
  \definecolor{coral4}{rgb}{0.545098,0.243137,0.184314}
  \definecolor{tomato1}{rgb}{1.000000,0.388235,0.278431}
  \definecolor{tomato2}{rgb}{0.933333,0.360784,0.258824}
  \definecolor{tomato3}{rgb}{0.803922,0.309804,0.223529}
  \definecolor{tomato4}{rgb}{0.545098,0.211765,0.149020}
  \definecolor{OrangeRed1}{rgb}{1.000000,0.270588,0.000000}
  \definecolor{OrangeRed2}{rgb}{0.933333,0.250980,0.000000}
  \definecolor{OrangeRed3}{rgb}{0.803922,0.215686,0.000000}
  \definecolor{OrangeRed4}{rgb}{0.545098,0.145098,0.000000}
  \definecolor{red1}{rgb}{1.000000,0.000000,0.000000}
  \definecolor{red2}{rgb}{0.933333,0.000000,0.000000}
  \definecolor{red3}{rgb}{0.803922,0.000000,0.000000}
  \definecolor{red4}{rgb}{0.545098,0.000000,0.000000}
  \definecolor{DeepPink1}{rgb}{1.000000,0.078431,0.576471}
  \definecolor{DeepPink2}{rgb}{0.933333,0.070588,0.537255}
  \definecolor{DeepPink3}{rgb}{0.803922,0.062745,0.462745}
  \definecolor{DeepPink4}{rgb}{0.545098,0.039216,0.313726}
  \definecolor{HotPink1}{rgb}{1.000000,0.431373,0.705882}
  \definecolor{HotPink2}{rgb}{0.933333,0.415686,0.654902}
  \definecolor{HotPink3}{rgb}{0.803922,0.376471,0.564706}
  \definecolor{HotPink4}{rgb}{0.545098,0.227451,0.384314}
  \definecolor{pink1}{rgb}{1.000000,0.709804,0.772549}
  \definecolor{pink2}{rgb}{0.933333,0.662745,0.721569}
  \definecolor{pink3}{rgb}{0.803922,0.568627,0.619608}
  \definecolor{pink4}{rgb}{0.545098,0.388235,0.423529}
  \definecolor{LightPink1}{rgb}{1.000000,0.682353,0.725490}
  \definecolor{LightPink2}{rgb}{0.933333,0.635294,0.678431}
  \definecolor{LightPink3}{rgb}{0.803922,0.549020,0.584314}
  \definecolor{LightPink4}{rgb}{0.545098,0.372549,0.396078}
  \definecolor{PaleVioletRed1}{rgb}{1.000000,0.509804,0.670588}
  \definecolor{PaleVioletRed2}{rgb}{0.933333,0.474510,0.623529}
  \definecolor{PaleVioletRed3}{rgb}{0.803922,0.407843,0.537255}
  \definecolor{PaleVioletRed4}{rgb}{0.545098,0.278431,0.364706}
  \definecolor{maroon1}{rgb}{1.000000,0.203922,0.701961}
  \definecolor{maroon2}{rgb}{0.933333,0.188235,0.654902}
  \definecolor{maroon3}{rgb}{0.803922,0.160784,0.564706}
  \definecolor{maroon4}{rgb}{0.545098,0.109804,0.384314}
  \definecolor{VioletRed1}{rgb}{1.000000,0.243137,0.588235}
  \definecolor{VioletRed2}{rgb}{0.933333,0.227451,0.549020}
  \definecolor{VioletRed3}{rgb}{0.803922,0.196078,0.470588}
  \definecolor{VioletRed4}{rgb}{0.545098,0.133333,0.321569}
  \definecolor{magenta1}{rgb}{1.000000,0.000000,1.000000}
  \definecolor{magenta2}{rgb}{0.933333,0.000000,0.933333}
  \definecolor{magenta3}{rgb}{0.803922,0.000000,0.803922}
  \definecolor{magenta4}{rgb}{0.545098,0.000000,0.545098}
  \definecolor{orchid1}{rgb}{1.000000,0.513726,0.980392}
  \definecolor{orchid2}{rgb}{0.933333,0.478431,0.913725}
  \definecolor{orchid3}{rgb}{0.803922,0.411765,0.788235}
  \definecolor{orchid4}{rgb}{0.545098,0.278431,0.537255}
  \definecolor{plum1}{rgb}{1.000000,0.733333,1.000000}
  \definecolor{plum2}{rgb}{0.933333,0.682353,0.933333}
  \definecolor{plum3}{rgb}{0.803922,0.588235,0.803922}
  \definecolor{plum4}{rgb}{0.545098,0.400000,0.545098}
  \definecolor{MediumOrchid1}{rgb}{0.878431,0.400000,1.000000}
  \definecolor{MediumOrchid2}{rgb}{0.819608,0.372549,0.933333}
  \definecolor{MediumOrchid3}{rgb}{0.705882,0.321569,0.803922}
  \definecolor{MediumOrchid4}{rgb}{0.478431,0.215686,0.545098}
  \definecolor{DarkOrchid1}{rgb}{0.749020,0.243137,1.000000}
  \definecolor{DarkOrchid2}{rgb}{0.698039,0.227451,0.933333}
  \definecolor{DarkOrchid3}{rgb}{0.603922,0.196078,0.803922}
  \definecolor{DarkOrchid4}{rgb}{0.407843,0.133333,0.545098}
  \definecolor{purple1}{rgb}{0.607843,0.188235,1.000000}
  \definecolor{purple2}{rgb}{0.568627,0.172549,0.933333}
  \definecolor{purple3}{rgb}{0.490196,0.149020,0.803922}
  \definecolor{purple4}{rgb}{0.333333,0.101961,0.545098}
  \definecolor{MediumPurple1}{rgb}{0.670588,0.509804,1.000000}
  \definecolor{MediumPurple2}{rgb}{0.623529,0.474510,0.933333}
  \definecolor{MediumPurple3}{rgb}{0.537255,0.407843,0.803922}
  \definecolor{MediumPurple4}{rgb}{0.364706,0.278431,0.545098}
  \definecolor{thistle1}{rgb}{1.000000,0.882353,1.000000}
  \definecolor{thistle2}{rgb}{0.933333,0.823529,0.933333}
  \definecolor{thistle3}{rgb}{0.803922,0.709804,0.803922}
  \definecolor{thistle4}{rgb}{0.545098,0.482353,0.545098}
  \definecolor{gray0}{rgb}{0.000000,0.000000,0.000000}
  \definecolor{grey0}{rgb}{0.000000,0.000000,0.000000}
  \definecolor{gray1}{rgb}{0.011765,0.011765,0.011765}
  \definecolor{grey1}{rgb}{0.011765,0.011765,0.011765}
  \definecolor{gray2}{rgb}{0.019608,0.019608,0.019608}
  \definecolor{grey2}{rgb}{0.019608,0.019608,0.019608}
  \definecolor{gray3}{rgb}{0.031373,0.031373,0.031373}
  \definecolor{grey3}{rgb}{0.031373,0.031373,0.031373}
  \definecolor{gray4}{rgb}{0.039216,0.039216,0.039216}
  \definecolor{grey4}{rgb}{0.039216,0.039216,0.039216}
  \definecolor{gray5}{rgb}{0.050980,0.050980,0.050980}
  \definecolor{grey5}{rgb}{0.050980,0.050980,0.050980}
  \definecolor{gray6}{rgb}{0.058824,0.058824,0.058824}
  \definecolor{grey6}{rgb}{0.058824,0.058824,0.058824}
  \definecolor{gray7}{rgb}{0.070588,0.070588,0.070588}
  \definecolor{grey7}{rgb}{0.070588,0.070588,0.070588}
  \definecolor{gray8}{rgb}{0.078431,0.078431,0.078431}
  \definecolor{grey8}{rgb}{0.078431,0.078431,0.078431}
  \definecolor{gray9}{rgb}{0.090196,0.090196,0.090196}
  \definecolor{grey9}{rgb}{0.090196,0.090196,0.090196}
  \definecolor{gray10}{rgb}{0.101961,0.101961,0.101961}
  \definecolor{grey10}{rgb}{0.101961,0.101961,0.101961}
  \definecolor{gray11}{rgb}{0.109804,0.109804,0.109804}
  \definecolor{grey11}{rgb}{0.109804,0.109804,0.109804}
  \definecolor{gray12}{rgb}{0.121569,0.121569,0.121569}
  \definecolor{grey12}{rgb}{0.121569,0.121569,0.121569}
  \definecolor{gray13}{rgb}{0.129412,0.129412,0.129412}
  \definecolor{grey13}{rgb}{0.129412,0.129412,0.129412}
  \definecolor{gray14}{rgb}{0.141176,0.141176,0.141176}
  \definecolor{grey14}{rgb}{0.141176,0.141176,0.141176}
  \definecolor{gray15}{rgb}{0.149020,0.149020,0.149020}
  \definecolor{grey15}{rgb}{0.149020,0.149020,0.149020}
  \definecolor{gray16}{rgb}{0.160784,0.160784,0.160784}
  \definecolor{grey16}{rgb}{0.160784,0.160784,0.160784}
  \definecolor{gray17}{rgb}{0.168627,0.168627,0.168627}
  \definecolor{grey17}{rgb}{0.168627,0.168627,0.168627}
  \definecolor{gray18}{rgb}{0.180392,0.180392,0.180392}
  \definecolor{grey18}{rgb}{0.180392,0.180392,0.180392}
  \definecolor{gray19}{rgb}{0.188235,0.188235,0.188235}
  \definecolor{grey19}{rgb}{0.188235,0.188235,0.188235}
  \definecolor{gray20}{rgb}{0.200000,0.200000,0.200000}
  \definecolor{grey20}{rgb}{0.200000,0.200000,0.200000}
  \definecolor{gray21}{rgb}{0.211765,0.211765,0.211765}
  \definecolor{grey21}{rgb}{0.211765,0.211765,0.211765}
  \definecolor{gray22}{rgb}{0.219608,0.219608,0.219608}
  \definecolor{grey22}{rgb}{0.219608,0.219608,0.219608}
  \definecolor{gray23}{rgb}{0.231373,0.231373,0.231373}
  \definecolor{grey23}{rgb}{0.231373,0.231373,0.231373}
  \definecolor{gray24}{rgb}{0.239216,0.239216,0.239216}
  \definecolor{grey24}{rgb}{0.239216,0.239216,0.239216}
  \definecolor{gray25}{rgb}{0.250980,0.250980,0.250980}
  \definecolor{grey25}{rgb}{0.250980,0.250980,0.250980}
  \definecolor{gray26}{rgb}{0.258824,0.258824,0.258824}
  \definecolor{grey26}{rgb}{0.258824,0.258824,0.258824}
  \definecolor{gray27}{rgb}{0.270588,0.270588,0.270588}
  \definecolor{grey27}{rgb}{0.270588,0.270588,0.270588}
  \definecolor{gray28}{rgb}{0.278431,0.278431,0.278431}
  \definecolor{grey28}{rgb}{0.278431,0.278431,0.278431}
  \definecolor{gray29}{rgb}{0.290196,0.290196,0.290196}
  \definecolor{grey29}{rgb}{0.290196,0.290196,0.290196}
  \definecolor{gray30}{rgb}{0.301961,0.301961,0.301961}
  \definecolor{grey30}{rgb}{0.301961,0.301961,0.301961}
  \definecolor{gray31}{rgb}{0.309804,0.309804,0.309804}
  \definecolor{grey31}{rgb}{0.309804,0.309804,0.309804}
  \definecolor{gray32}{rgb}{0.321569,0.321569,0.321569}
  \definecolor{grey32}{rgb}{0.321569,0.321569,0.321569}
  \definecolor{gray33}{rgb}{0.329412,0.329412,0.329412}
  \definecolor{grey33}{rgb}{0.329412,0.329412,0.329412}
  \definecolor{gray34}{rgb}{0.341176,0.341176,0.341176}
  \definecolor{grey34}{rgb}{0.341176,0.341176,0.341176}
  \definecolor{gray35}{rgb}{0.349020,0.349020,0.349020}
  \definecolor{grey35}{rgb}{0.349020,0.349020,0.349020}
  \definecolor{gray36}{rgb}{0.360784,0.360784,0.360784}
  \definecolor{grey36}{rgb}{0.360784,0.360784,0.360784}
  \definecolor{gray37}{rgb}{0.368627,0.368627,0.368627}
  \definecolor{grey37}{rgb}{0.368627,0.368627,0.368627}
  \definecolor{gray38}{rgb}{0.380392,0.380392,0.380392}
  \definecolor{grey38}{rgb}{0.380392,0.380392,0.380392}
  \definecolor{gray39}{rgb}{0.388235,0.388235,0.388235}
  \definecolor{grey39}{rgb}{0.388235,0.388235,0.388235}
  \definecolor{gray40}{rgb}{0.400000,0.400000,0.400000}
  \definecolor{grey40}{rgb}{0.400000,0.400000,0.400000}
  \definecolor{gray41}{rgb}{0.411765,0.411765,0.411765}
  \definecolor{grey41}{rgb}{0.411765,0.411765,0.411765}
  \definecolor{gray42}{rgb}{0.419608,0.419608,0.419608}
  \definecolor{grey42}{rgb}{0.419608,0.419608,0.419608}
  \definecolor{gray43}{rgb}{0.431373,0.431373,0.431373}
  \definecolor{grey43}{rgb}{0.431373,0.431373,0.431373}
  \definecolor{gray44}{rgb}{0.439216,0.439216,0.439216}
  \definecolor{grey44}{rgb}{0.439216,0.439216,0.439216}
  \definecolor{gray45}{rgb}{0.450980,0.450980,0.450980}
  \definecolor{grey45}{rgb}{0.450980,0.450980,0.450980}
  \definecolor{gray46}{rgb}{0.458824,0.458824,0.458824}
  \definecolor{grey46}{rgb}{0.458824,0.458824,0.458824}
  \definecolor{gray47}{rgb}{0.470588,0.470588,0.470588}
  \definecolor{grey47}{rgb}{0.470588,0.470588,0.470588}
  \definecolor{gray48}{rgb}{0.478431,0.478431,0.478431}
  \definecolor{grey48}{rgb}{0.478431,0.478431,0.478431}
  \definecolor{gray49}{rgb}{0.490196,0.490196,0.490196}
  \definecolor{grey49}{rgb}{0.490196,0.490196,0.490196}
  \definecolor{gray50}{rgb}{0.498039,0.498039,0.498039}
  \definecolor{grey50}{rgb}{0.498039,0.498039,0.498039}
  \definecolor{gray51}{rgb}{0.509804,0.509804,0.509804}
  \definecolor{grey51}{rgb}{0.509804,0.509804,0.509804}
  \definecolor{gray52}{rgb}{0.521569,0.521569,0.521569}
  \definecolor{grey52}{rgb}{0.521569,0.521569,0.521569}
  \definecolor{gray53}{rgb}{0.529412,0.529412,0.529412}
  \definecolor{grey53}{rgb}{0.529412,0.529412,0.529412}
  \definecolor{gray54}{rgb}{0.541176,0.541176,0.541176}
  \definecolor{grey54}{rgb}{0.541176,0.541176,0.541176}
  \definecolor{gray55}{rgb}{0.549020,0.549020,0.549020}
  \definecolor{grey55}{rgb}{0.549020,0.549020,0.549020}
  \definecolor{gray56}{rgb}{0.560784,0.560784,0.560784}
  \definecolor{grey56}{rgb}{0.560784,0.560784,0.560784}
  \definecolor{gray57}{rgb}{0.568627,0.568627,0.568627}
  \definecolor{grey57}{rgb}{0.568627,0.568627,0.568627}
  \definecolor{gray58}{rgb}{0.580392,0.580392,0.580392}
  \definecolor{grey58}{rgb}{0.580392,0.580392,0.580392}
  \definecolor{gray59}{rgb}{0.588235,0.588235,0.588235}
  \definecolor{grey59}{rgb}{0.588235,0.588235,0.588235}
  \definecolor{gray60}{rgb}{0.600000,0.600000,0.600000}
  \definecolor{grey60}{rgb}{0.600000,0.600000,0.600000}
  \definecolor{gray61}{rgb}{0.611765,0.611765,0.611765}
  \definecolor{grey61}{rgb}{0.611765,0.611765,0.611765}
  \definecolor{gray62}{rgb}{0.619608,0.619608,0.619608}
  \definecolor{grey62}{rgb}{0.619608,0.619608,0.619608}
  \definecolor{gray63}{rgb}{0.631373,0.631373,0.631373}
  \definecolor{grey63}{rgb}{0.631373,0.631373,0.631373}
  \definecolor{gray64}{rgb}{0.639216,0.639216,0.639216}
  \definecolor{grey64}{rgb}{0.639216,0.639216,0.639216}
  \definecolor{gray65}{rgb}{0.650980,0.650980,0.650980}
  \definecolor{grey65}{rgb}{0.650980,0.650980,0.650980}
  \definecolor{gray66}{rgb}{0.658824,0.658824,0.658824}
  \definecolor{grey66}{rgb}{0.658824,0.658824,0.658824}
  \definecolor{gray67}{rgb}{0.670588,0.670588,0.670588}
  \definecolor{grey67}{rgb}{0.670588,0.670588,0.670588}
  \definecolor{gray68}{rgb}{0.678431,0.678431,0.678431}
  \definecolor{grey68}{rgb}{0.678431,0.678431,0.678431}
  \definecolor{gray69}{rgb}{0.690196,0.690196,0.690196}
  \definecolor{grey69}{rgb}{0.690196,0.690196,0.690196}
  \definecolor{gray70}{rgb}{0.701961,0.701961,0.701961}
  \definecolor{grey70}{rgb}{0.701961,0.701961,0.701961}
  \definecolor{gray71}{rgb}{0.709804,0.709804,0.709804}
  \definecolor{grey71}{rgb}{0.709804,0.709804,0.709804}
  \definecolor{gray72}{rgb}{0.721569,0.721569,0.721569}
  \definecolor{grey72}{rgb}{0.721569,0.721569,0.721569}
  \definecolor{gray73}{rgb}{0.729412,0.729412,0.729412}
  \definecolor{grey73}{rgb}{0.729412,0.729412,0.729412}
  \definecolor{gray74}{rgb}{0.741176,0.741176,0.741176}
  \definecolor{grey74}{rgb}{0.741176,0.741176,0.741176}
  \definecolor{gray75}{rgb}{0.749020,0.749020,0.749020}
  \definecolor{grey75}{rgb}{0.749020,0.749020,0.749020}
  \definecolor{gray76}{rgb}{0.760784,0.760784,0.760784}
  \definecolor{grey76}{rgb}{0.760784,0.760784,0.760784}
  \definecolor{gray77}{rgb}{0.768627,0.768627,0.768627}
  \definecolor{grey77}{rgb}{0.768627,0.768627,0.768627}
  \definecolor{gray78}{rgb}{0.780392,0.780392,0.780392}
  \definecolor{grey78}{rgb}{0.780392,0.780392,0.780392}
  \definecolor{gray79}{rgb}{0.788235,0.788235,0.788235}
  \definecolor{grey79}{rgb}{0.788235,0.788235,0.788235}
  \definecolor{gray80}{rgb}{0.800000,0.800000,0.800000}
  \definecolor{grey80}{rgb}{0.800000,0.800000,0.800000}
  \definecolor{gray81}{rgb}{0.811765,0.811765,0.811765}
  \definecolor{grey81}{rgb}{0.811765,0.811765,0.811765}
  \definecolor{gray82}{rgb}{0.819608,0.819608,0.819608}
  \definecolor{grey82}{rgb}{0.819608,0.819608,0.819608}
  \definecolor{gray83}{rgb}{0.831373,0.831373,0.831373}
  \definecolor{grey83}{rgb}{0.831373,0.831373,0.831373}
  \definecolor{gray84}{rgb}{0.839216,0.839216,0.839216}
  \definecolor{grey84}{rgb}{0.839216,0.839216,0.839216}
  \definecolor{gray85}{rgb}{0.850980,0.850980,0.850980}
  \definecolor{grey85}{rgb}{0.850980,0.850980,0.850980}
  \definecolor{gray86}{rgb}{0.858824,0.858824,0.858824}
  \definecolor{grey86}{rgb}{0.858824,0.858824,0.858824}
  \definecolor{gray87}{rgb}{0.870588,0.870588,0.870588}
  \definecolor{grey87}{rgb}{0.870588,0.870588,0.870588}
  \definecolor{gray88}{rgb}{0.878431,0.878431,0.878431}
  \definecolor{grey88}{rgb}{0.878431,0.878431,0.878431}
  \definecolor{gray89}{rgb}{0.890196,0.890196,0.890196}
  \definecolor{grey89}{rgb}{0.890196,0.890196,0.890196}
  \definecolor{gray90}{rgb}{0.898039,0.898039,0.898039}
  \definecolor{grey90}{rgb}{0.898039,0.898039,0.898039}
  \definecolor{gray91}{rgb}{0.909804,0.909804,0.909804}
  \definecolor{grey91}{rgb}{0.909804,0.909804,0.909804}
  \definecolor{gray92}{rgb}{0.921569,0.921569,0.921569}
  \definecolor{grey92}{rgb}{0.921569,0.921569,0.921569}
  \definecolor{gray93}{rgb}{0.929412,0.929412,0.929412}
  \definecolor{grey93}{rgb}{0.929412,0.929412,0.929412}
  \definecolor{gray94}{rgb}{0.941176,0.941176,0.941176}
  \definecolor{grey94}{rgb}{0.941176,0.941176,0.941176}
  \definecolor{gray95}{rgb}{0.949020,0.949020,0.949020}
  \definecolor{grey95}{rgb}{0.949020,0.949020,0.949020}
  \definecolor{gray96}{rgb}{0.960784,0.960784,0.960784}
  \definecolor{grey96}{rgb}{0.960784,0.960784,0.960784}
  \definecolor{gray97}{rgb}{0.968627,0.968627,0.968627}
  \definecolor{grey97}{rgb}{0.968627,0.968627,0.968627}
  \definecolor{gray98}{rgb}{0.980392,0.980392,0.980392}
  \definecolor{grey98}{rgb}{0.980392,0.980392,0.980392}
  \definecolor{gray99}{rgb}{0.988235,0.988235,0.988235}
  \definecolor{grey99}{rgb}{0.988235,0.988235,0.988235}
  \definecolor{gray100}{rgb}{1.000000,1.000000,1.000000}
  \definecolor{grey100}{rgb}{1.000000,1.000000,1.000000}
  \definecolor{dark grey}{rgb}{0.662745,0.662745,0.662745}
  \definecolor{DarkGrey}{rgb}{0.662745,0.662745,0.662745}
  \definecolor{dark gray}{rgb}{0.662745,0.662745,0.662745}
  \definecolor{DarkGray}{rgb}{0.662745,0.662745,0.662745}
  \definecolor{dark blue}{rgb}{0.000000,0.000000,0.545098}
  \definecolor{DarkBlue}{rgb}{0.000000,0.000000,0.545098}
  \definecolor{dark cyan}{rgb}{0.000000,0.545098,0.545098}
  \definecolor{DarkCyan}{rgb}{0.000000,0.545098,0.545098}
  \definecolor{dark magenta}{rgb}{0.545098,0.000000,0.545098}
  \definecolor{DarkMagenta}{rgb}{0.545098,0.000000,0.545098}
  \definecolor{dark red}{rgb}{0.545098,0.000000,0.000000}
  \definecolor{DarkRed}{rgb}{0.545098,0.000000,0.000000}
  \definecolor{light green}{rgb}{0.564706,0.933333,0.564706}
  \definecolor{LightGreen}{rgb}{0.564706,0.933333,0.564706}

\RequirePackage[colorlinks,citecolor=blue, linkcolor=blue, urlcolor=blue]{hyperref}
\newtheorem{prop} {Proposition}

\newtheorem{cor} {Corollary}

\newtheorem{condition} {Condition}

\usepackage{xr}
\begin{document}
\title{\vspace{-7ex} Extremal Depth for Functional Data and Applications  }
\author{Naveen N. Narisetty and Vijayan N. Nair  \footnote{Naveen N. Narisetty is  PhD Candidate, Department of Statistics, University of Michigan, Ann Arbor, MI 48109
(email: naveennn@umich.edu); and Vijayan N. Nair is Professor, Department of Statistics and Department of Industrial \& Operations Engineering, University of Michigan, Ann Arbor, MI 48109 (email: vnn@umich.edu).  The authors are grateful to the editors and the referees for their comments and suggestions that have led to significant improvements of an earlier version of this article. The article is accepted for publication by Journal of the American Statistical Association (Theory and Methods).}}
\date{\vspace{-7ex}} 
\maketitle
\begin{abstract}
We propose a new notion called `extremal depth' (ED) for functional data, discuss its properties, and compare its performance with existing concepts. The proposed notion is based on a measure of extreme `outlyingness'.  ED has several desirable properties that are not shared by other notions and is especially well suited for obtaining central regions of functional data and function spaces. In particular: a) the central region achieves the nominal (desired) simultaneous coverage probability; b) there is a correspondence between ED-based (simultaneous) central regions and appropriate pointwise central regions; and c) the method is resistant to certain classes of functional outliers. The paper examines the performance of ED and compares it with other depth notions. Its usefulness is demonstrated through applications to constructing central regions, functional boxplots, outlier detection, and simultaneous confidence bands in regression problems.\\

\noindent \emph{Keywords:} data depth, central regions, functional boxplots, outlier detection, simultaneous inference \vspace{0.4in}
\end{abstract}

\section{Introduction}
Ranks, order-statistics, and quantiles have been used extensively  for statistical inference with univariate data.  Many authors have studied their generalizations for multivariate data using notions of ``data depth". The classical measure based on Mahalanobis distance \citep{Mahalanobis36} is ideally suited for multivariate normal (or more generally elliptical) distributions. Tukey's half-space depth \citep{Tukey75} appears to be the first new notion for the multivariate case, and there has been a lot of work since then. \cite{Brown83} defined a `median' for multivariate data using the $L_1$ metric, and \cite{Vardi00} extended this to obtain a notion of multivariate depth. Other concepts include simplicial depth \citep{Liu90}, geometric notion of quantiles \citep{Chaudhuri96}, projection depth  \citep{Zuo00, Zuo03}, and  spatial depth  \citep{Vardi00, Serfling02}.  See \cite{Zuo00} for a  review. Various types of statistical inference have also been based on multivariate depth notions, including classification \citep{Jornsten04, Ghosh05,Li12}, outlier detection \citep{Donoho92, Mosler02}, and hypothesis testing \citep{Liu97}.  \cite{Liu99} studied the use of depth-based  methods for inference on distributional quantities such as location, scale, bias, skewness and kurtosis.

{ For functional data, }\cite{Fraiman01} proposed integrated data depth (ID); \cite{Lopez09} introduced band depth (BD) and modified band depth (MBD); and \cite{Lopez11} proposed a half-region depth (HRD). Several other notions of depth for multivariate data have also been extended to functional data. For instance, \cite{Chakraborty12} developed spatial depth (SD) for functional data. One can also extend \cite{Zuo03}'s projection-based depth functions and multivariate medians to functional data. However, several of these notions and extensions suffer from a ``degeneracy" problem pointed out in \cite{Chakraborty12}. Specifically, in infinite-dimensional function spaces, with probability one, all the functions will have zero depth  \citep{Chakraborty12, Chakraborty14}.

As with multivariate data, functional depth can be used for many applications. \cite{Fraiman01} used ID for constructing trimmed functional mean. \cite{Lopez06} used BD for classification of functional data,  \cite{Sun11} proposed functional boxplots based on MBD, {  \cite{Pereira14} proposed some notions of extremality for functional data and used them to construct rank tests,} and \cite{Hubert15} considered functional outlier detection based on some measures of depth and outlyingness. Depth notions can also be used to obtain central regions of data which, for instance, form the basis for constructing boxplots.

{Both ID and MBD are based on some form of averaging of the depth at different points }in the domain and, as a result, their depth level sets are not convex. This has important implications for corresponding central regions as discussed in later sections. In addition, they may not be resistant to functions that are outlying in small regions of the domain.

This paper develops a new notion called Extremal Depth (ED) for functional data. We will show that ED and associated central regions possess several attractive features including: \vspace{-2ex}
\begin{itemize}
\item ED central regions achieve their nominal coverage \emph{exactly} due to the convexity of the depth contours; \vspace{-2ex}
\item There is a direct correspondence between the (simultaneous) ED central regions and the usual pointwise central regions based on
quantiles; as a consequence, the width of the ED simultaneous central regions is, roughly speaking, proportional to a measure of variation at each point; and \vspace{-2ex}
\item ED central regions are resistant to functions that are `outlying' even in a small region of the domain. \vspace{-2ex}
\end{itemize}
These features lead to desirable properties for corresponding functional boxplots, simultaneous confidence regions for function estimation, and outlier detection. 

The rest of the article is organized as follows.  Section \ref{ed-sec} introduces ED for a sample of functional data and illustrates it on a real dataset. Section \ref{edgen-sec} defines ED for general probability distributions and discusses its theoretical properties.  Section \ref{central-sec} deals with construction of central regions of functional data and develops several results including exact coverage and correspondence to pointwise regions. Section \ref{fbplot-sec} describes applications to functional boxplots and outlier detection, and the advantages of ED-based methods over others. Section \ref{more-apps} demonstrates how ED can be used to construct simultaneous confidence bands for functional parameters. \vspace{-2ex}

\section{Extremal Depth} \label{ed-sec}  \vspace{-2ex}
\subsection{Depth distribution}
Let $S:=\{ f_1(t), f_2(t), \cdots, f_n(t)\}$ be a collection of $n$ functional observations with $t \in  {\Tau}$. For ease of exposition, we assume throughout that the functions are continuous and infinite-dimensional and, without loss of generality, we take the domain $\mathscr{\Tau}$ to be $[0, 1]$. However, as with other notions, ED can also be used for functional observations observed at a finite number of points.

Let $g(t)$ be a given function that may or may not be a member of $S$. For each fixed $t \in [0, 1]$, define the pointwise depth of $g(t)$ with respect to $S$ as \vspace{-2ex}
\begin{equation}\label{pointwise-depth}
\begin{array}{rll}
\displaystyle D_{g} (t, S) &  \displaystyle :=  1 - \frac{|\sum_{i=1}^n [\mathbbm{1}\{ f_i(t) < g(t) \} -  \mathbbm{1}\{ f_i(t) > g(t) \}] |}{n}. \vspace{-2ex}
\end{array}
\end{equation}
Thus, any given function $g(\cdot)$ is mapped into the pointwise depth function $D_g(\cdot, S)$ whose range is $\mathbb{D}_g \subset\{0, 1/n, 2/n, \cdots, 1 \}$. Let $\mathbb{D}$ be the union of $\mathbb{D}_g $ over all functions $g$. We call $\mathbb{D}$ the \emph{set of depth values}.

Let $\Phi_g(\cdot)$ be the cumulative distributions function (CDF) of the distinct values taken by $D_g(t, S)$ as $t$ varies in $[0, ~1]$. This will be called the depth CDF or d-CDF and defined formally as \vspace{-2ex}
\begin{equation}\label{dcdf}
 \Phi_g (r) = \int_{0}^1  \mathbbm{1}\{D_{g} (t, S) \leq r\}dt , \vspace{-2ex}
\end{equation}
for each fixed $r \in \mathbb{D}$.  Note that if $ \Phi_g $ has most of its mass close to zero (or one), then $g$ is away from (or close to) the center of the data. (See the illustrative example in Figure \ref{nocross-fig} for computation of d-CDFs.)

{   In this paper, we will focus on depth measures based on d-CDFs  (distributions). One concern might be that the d-CDFs do not use the dependence structure of the functional data. However, such information is generally unavailable. Therefore, we will develop methods that do not require the knowledge of the dependence relationships and that can be applied to a broad class of problems. Note that this is also the approach taken in the literature on functional and multivariate depth.}

We need an appropriate way to order these d-CDFs to get a one-dimensional notion of depth. (Clearly, there is no single approach that will dominate all others, so one has to decide on the appropriate one by examining its performance under different situations.) First-order stochastic dominance may appear to be the most natural way to order distributions, but it is not useful here except in the trivial case where the functions do not cross. Alternatively, one can use a simple functional of the d-CDFs such as the mean or median. {   In fact, the integrated depth (or ID)  by \cite{Fraiman01} corresponds (approximately) to  the average pointwise depth $\int_0^1 D_g(t, S) ~dt,$ which is also the mean corresponding to the depth distribution $\Phi_g (\cdot)$. [We say ``approximately'' due to the minor difference: we use $[\mathbbm{1}\{ f_i(t) < g(t) \} -  \mathbbm{1}\{ f_i(t) > g(t) \}]$  in the definition of $D_g$ while \cite{Fraiman01}'s definition of ID is based on $[\mathbbm{1}\{ f_i(t) \leq g(t) \} -  \mathbbm{1}\{ f_i(t) > g(t) \}]$.] Modified Band Depth (MBD) is also related to the d-CDFs although the relationship between d-CDFs and MBD is more complex than the one between d-CDFs and ID. This is because MBD can be expressed as the average of the univariate simplicial depth, and the univariate simplicial depth has a monotone relationship with the univariate depth $D_g(t,S)$. Resultantly, MBD corresponds to the mean of a non-trivial function with respect to the depth distribution $\Phi_g (\cdot)$.} We will provide a comparison of various functional depths in Section \ref{edgen-sec}.


\subsection{Definition of Extremal Depth}

Our notion of extremal depth will be based on a comparison of $\Phi_g(r)$, the d-CDFs,  for $r$ near zero. It focuses on the left tail of the distribution and can be viewed as left-tail stochastic ordering. The idea can be explained simply as follows. Consider two functions $g$ and $h$ with corresponding d-CDFs $\Phi_g$ and $\Phi_h$. Let $0 \leq d_1< d_2< \cdots < d_M \leq 1$ be the ordered elements of their combined depth levels. If $\Phi_h(d_1) > \Phi_g(d_1)$, then $h \prec g$ (or equivalently $g \succ h$, and is read as $h$ is more extreme then $g$); if $\Phi_g(d_1) > \Phi_h(d_1)$, then $ h \succ g$. If $\Phi_g(d_1) = \Phi_h(d_1)$, we move to $d_2$ and make a similar comparison based on their values at $d_2$. The comparison is repeated until the tie is broken. If $\Phi_g(d_i) = \Phi_h(d_i)$ for all $i = 1, ... M$, the two functions are equivalent in terms of depth and are denoted as $g \sim h$. (This ordering is defined formally in Section \ref{edgen-sec} when we consider a more general context with arbitrary function spaces $S$ and distributions.)

The extremal depth (ED) of a function $g$ with respect to the sample $S = \{f_1, \cdots, f_n \}$ can now be defined as \vspace{-1ex}
\begin{equation}\label{depth-dpmf}
ED(g, S) = \frac{ \# \{i : g \succeq f_i \}}{n}, \vspace{-2ex}
\end{equation}
where $ g \succeq f_i $ if either $ g \succ f_i $ or $ g \sim f_i$. If $g \in S$, then this is just the normalized rank of $g$; i.e., $ED(g, S) = R(g, S) / n$ where $R(g, S) = \{i: g \succeq f_i \}$ is the rank of $g$. This relationship between ED and its rank is similar to corresponding relationships of normalized rank functions for some other depth notions in the literature \citep{Liu93, Lopez09}. The distinguishing feature of ED is the nature of the ordering, i.e., left-tail stochastic ordering of the depth distributions.

The ED median of a set of functional observations $S$ can be defined (in an obvious manner) as the function (or functions) in $S$ that has (or have) the largest depth. ED median also has the following { max-min interpretation}. For a function $g \in S$,
let $d_{\min}(g) = \inf_{t \in [0, 1]} D_{g}(t, S)$, the pointwise depth in Equation (\ref{pointwise-depth}). Then, if $g$ is an ED median, $d_{\min}(g)$ attains the maximum: $ d_{\min}(g) = \max_{1 \le k \le n} d_{\min}(f_k)$; i.e., an ED median maximizes the minimum pointwise depth over $t \in [0, ~1]$. {   This is so because otherwise, there exists a sample function $f_j$ having $d_{\min}(f_j)$ larger than $d_{\min}(g)$. This implies that $\Phi_{f_j} (d_{\min}(g))= 0$ but $ \Phi_{g} (d_{\min}(g)) >0$. The definition of ED then implies that $g \prec f_j$ and  $ED(g, S) < ED(f_j, S)$, which is a contradiction because $g$ is a median.}

\begin{figure}[h]
\caption{An illustrative example: (a) eight sample functions and (b) their depth CDF's. The columns correspond to each of four depth levels $\{1/8, 3/8, 5/8,7/8 \}$ and the rows correspond to different sample functions. \label{nocross-fig} }
\centering
\includegraphics[width = 6.4in,height=2.8in]{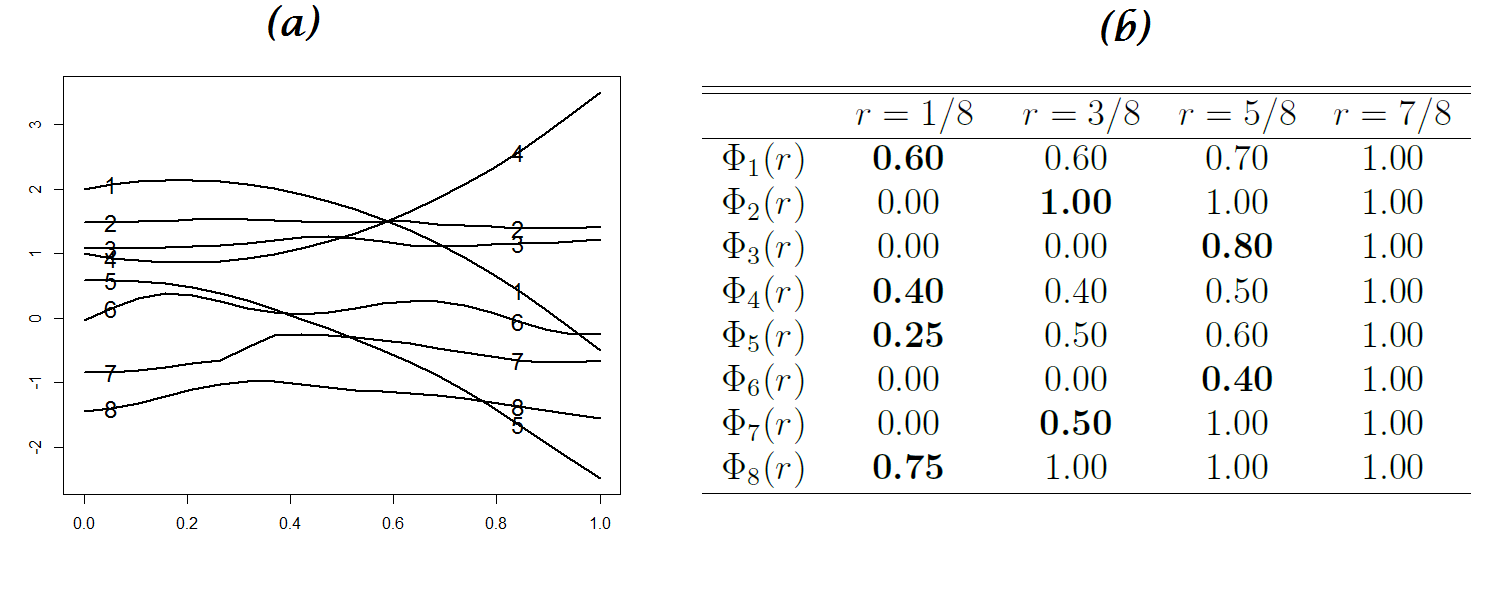} \vspace{-7ex}
\end{figure}
We now consider the illustrative example in  Figure \ref{nocross-fig} (a) with eight sample functions.  The d-CDF's of all the functions  are shown as a table in Figure \ref{nocross-fig} (b). ED gives the ordering $ { f_8} \prec { f_1} \prec   f_4  \prec  f_5  \prec   f_2  \prec  f_7  \prec  f_3  \prec  f_6.$ So $f_8$ is the most extreme observation and $f_6$ is the deepest (median). Note that the ordering $f_8 \prec f_1 \prec f_4 \prec f_5$ is based on a comparison of the d-CDF values at $r = 1/8$ (these values are in bold); the ordering $f_2 \prec f_7$ is based on their d-CDF values at $r = 3/8$; and the ordering $f_3 \prec f_6$ is based on their values at $r = 5/8$. From this, we get the extremal depths of these functions as: $ED(f_8) = 1/8, ED(f_1) = 2/8$ and so on.

\begin{figure}[h]
\caption{Orthosis data example: The three panels show the 240 functional observations (in gray) along with their two most outlying functions (in red) and the median (in blue) using ED, ID and MBD, respectively.  \label{figex1}}
\centering
\includegraphics[width = 6.3in, height = 2.5in]{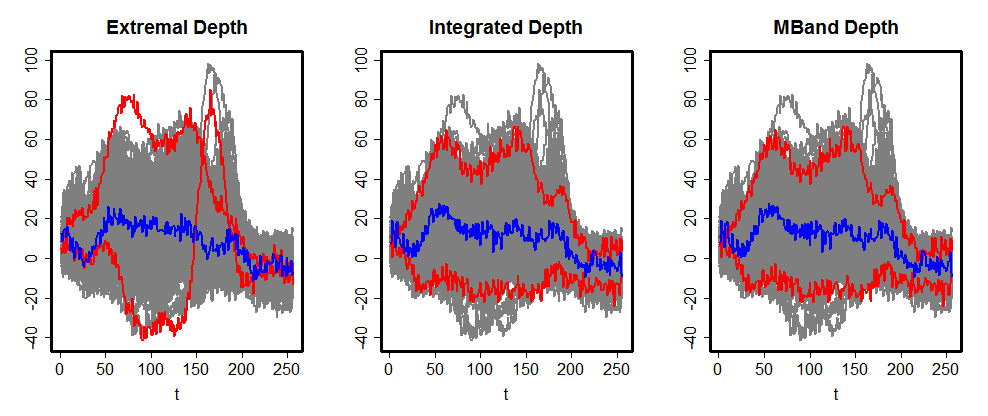} \vspace{-5ex}
\end{figure}

We now use the orthosis dataset  \citep{Cahouet02} to illustrate ED and visually compare its performance with ID and MBD. This dataset consists of moment of force measured at the knee under four different experimental conditions, measured at 256 equally-spaced time points for seven subjects with ten replications per subject. Figure \ref{figex1} shows the results for 240 functional observations from six subjects who have similar range of moment of force values. The x-axis represents time when the measurement is taken and the y-axis shows the resultant moment of force at the knee. The sample functions are plotted in gray, while the deepest function is in blue and the two least deep functions are in red.

The three panels in Figure \ref{figex1} correspond to ED, ID and MBD respectively. We restrict attention to ID and MBD in our empirical comparisons because these notions are commonly used, non-degenerate and invariant to monotone-transformations. (These properties are discussed in Section \ref{prop-sec}).  The medians for all three notions are qualitatively similar. However, the two extreme functions based on ID and MBD fall well within the boundaries of the entire data cloud while the two for ED are most extreme in at least some part of the domain. As we shall see, this is due to the non-convexity of the depth level sets of ID and MBD. { This example also illustrates that as ED is based on ``extreme outlyingness", it will penalize functions that are outliers in a short interval even if they are ``representative" in the rest of the domain. So, if one is particularly interested in characterizing the overall behavior of the functions, other measures of depth may be preferable. }\\

\section{ED for Theoretical (Population) Distributions and Its Properties} \label{edgen-sec}

There has been discussion of the desirable properties for depth notions in the literature \citep{ Liu90, Zuo00, Mosler12}. We will examine the performance of ED with respect to these properties and compare it with existing notions. To do this, we first have to extend  the notion of ED from sample data to theoretical (population) distributions.   \vspace{-2ex}

\subsection{Definition}

 Let $\mathbb{P}$ be a distribution on $C[0,1]$ and $X \sim \mathbb{P}$ be a random function. We denote $F_t$ to be the CDF of the random variable $X(t)$, and $\bar{F}_t (\cdot) = 1 - F_t(\cdot)$.  For any function $g$, define the depth of $g$ at $t$  as \vspace{-2ex}
\begin{equation}
\begin{array}{rll}
D_g(t, X) &:=& 1 -  \left|\mathbb{P} \left[ X(t) > g(t) \right] -  \mathbb{P} \left[ X(t) < g(t) \right]  \right|\\
 & =& 1 -  |\bar{F}_t(g(t)) - F_t (g(t)-) |. \vspace{-2ex}
\end{array}
\end{equation}
When the univariate distributions $F_t$ are continuous, $D_g(t, X)  =  1 -  |1 - 2 F_t (g(t)) |$, {   and is monotonically related to univariate half-space depth and simplicial depth, which are given by $\min (F_t (\cdot), 1- F_t (\cdot)),$ and $F_t (\cdot) (1 - F_t (\cdot)),$ respectively.}

The d-CDF of the function $g$ is defined, similar to the finite-sample case, as \vspace{-2ex}
\begin{equation}\label{depth-dist}
 \Phi_g (r) = \int_{[0,1]}  \mathbbm{1}\{D_g(t, X) \leq r\} \, d t, \vspace{-2ex}
\end{equation}
for $r \in [0,1]$. Note that, if necessary, one can replace the uniform weight distribution in the definition of $\Phi_g (r) $ by a weighted measure to give higher or lower importance to certain regions of the domain. 

As in the finite-sample case, we use the d-CDFs to obtain an ordering of functions. Because the d-CDFs now can be continuous, we need a slightly more general definition. Consider a pair of functions $g, h \in C[0,1]$, and define
\begin{equation}\label{dijplus}
r^* =   \inf \{r \in [0,1]: \Phi_{g}(r)  \neq \Phi_{h}(r) \}, \vspace{-2ex}
\end{equation}
the infimum of values at which d-CDFs of $g$ and $h$ differ. Then, we say $h \prec g$ ($h$ more extreme than $g$) if there exists $\delta >0$ such that $\Phi_{h}(r)  > \Phi_g(r)$ for all $r \in (r^*, ~r^*  + \delta).$ If $r^* <1,$ such a $\delta$ exists as long as $ \Phi_{g}$ and $ \Phi_{h}$ have finitely many crossings {   (see Appendix for a more general definition when such $\delta$ may not exist)}. If $r^* = 1,$ we say that $g \sim h$.

Extremal depth of a function $g$ w.r.t. the distribution $\mathbb{P}$ is now defined as \vspace{-2ex}
\begin{equation}\label{depth-def}
ED (g, \mathbb{P}) :=1 - \mathbb{P} \left[ g \prec X \right] = \mathbb{P} \left[ g \succeq X \right], \text{ where } X \sim \mathbb{P}. 
\end{equation}

\subsection{Properties}\label{prop-sec}
\cite{Liu90,Zuo00}  proposed several desirable properties for multivariate depth notions, and \cite{Mosler12} extended them for functional depth. The first four properties below are satisfied by ED, ID, BD and MBD but not by some others. The next two concepts discussed below, convexity and `null at the boundary' (NAB), are satisfied by ED but not by ID and MBD. The convexity property leads to desirable shapes for central regions as shown in the next section. The NAB property is also important and is related to being resistant to outliers. \\

\noindent {\bf Transitivity} (if $f_1 \prec f_2$ and $f_2 \prec f_3$, then $f_1 \prec f_3$) and {\bf invariance} under monotone transformations (order preserving as well as order reversing) are two well-known properties. It can be easily shown that ED satisfies them, as do ID, BD and MBD (where the ordering $f_1 \prec f_2$ for ID and MBD is interpreted as $f_2$ deeper than $f_1$). However, spatial depth (SD) \citep{Chakraborty12} does not satisfy the invariance property. The details are omitted. \\

{ 
\noindent{\bf Maximality of the center} {property requires that if there exists a natural center for the distribution of interest such as a center of symmetry, then it should have the highest depth. A center of symmetry can be formally defined as  $s \in C[0,1]$ satisfying: $\mathbb{P} \left[ X(t) > s(t) \right] =  \mathbb{P} \left[ X(t) < s(t) \right], \forall t \in [0,1] $.  ED has a depth of one at the center of symmetry if it exists. More generally, consider the following set
\begin{equation} \label{med}
\mathscr{M}:= \left \{m \in C[0,1] :  m(t) = \argmin \limits_{y} \left|\mathbb{P} \left[ X(t) > y \right] -  \mathbb{P} \left[ X(t) < y \right]  \right| \right \}.
\end{equation}
 When $\mathscr{M}$ is nonempty, any function $m \in \mathscr{M}$ has extremal depth equal to one because $m \succeq f, ~ \forall ~ f \in C[0,1]$. This follows easily because at any point $t \in [0,1]$, $D_t(m, X) \geq D_t(f, X), \forall f$. Note that $\mathscr{M}$ contains center of symmetry when it exists and generalizes center of symmetry when it doesn't exist. However, $\mathscr{M}$ is not guaranteed to be non-empty only in the irregular case that none of the functions satisfying the ``argmin condition'' in Equation \eqref{med} are continuous.

When a center of symmetry exists, ID and MBD also have it as the median but it may not necessarily be the median for SD. However, under the stronger notion of symmetry that $X - s$ and $s - X$ have the same distribution, SD has $s$ as the median. While BD also has the center of symmetry as its median, \cite{Chakraborty142} showed that, for many common stochastic processes, BD assigns a depth of zero to the center of symmetry, making it not deeper than any other function.\\ } }

\noindent {\bf Monotonicity from the center} requires that if $m$ is a median and functions $f$, $g$ are such that $\forall t$, either $m(t) \leq g(t) \leq f(t)$ or $m(t) \geq g(t) \geq f(t),$ then we require $g$ to be at least as deep as $f$.
For any median $m \in \mathscr{M}$, the monotonicity from the median property is satisfied for ED; the extremal depth does not increase as we move away from $m$. ED, ID, BD and MBD all satisfy monotonicity from the center of symmetry, when it exists. The proof is omitted. \\

\noindent{\bf Convex depth level sets:}
For a  given function $h$ and fixed $\alpha \in (0, ~1)$,  define the ED level set as $\{h: ED(h, \mathbb{P}) \geq  \alpha  \}.$
\begin{prop} \label{prop1}
Under a mild condition (Condition \ref{eq-cond}(b) in Appendix), the  ED level sets are convex for each $\alpha \in (0,1)$.
\end{prop}

This property is highly desirable for constructing central regions of a desired coverage $(1 - \alpha)$ (developed in the next section). Neither ID nor MBD is guaranteed to have convex depth level sets, which was already suggested by Figure \ref{figex1}. The proof is provided in the Appendix. \\

\noindent {\bf Null at the Boundary:} \citep{Mosler12} considered a depth notion to satisfy the `null at infinity' (NAI) property if $D(h, \mathbb{P}) \rightarrow 0$ as $\| h\| \rightarrow \infty$. It is shown in the Appendix that ED satisfies the NAI property. Neither ID nor MBD satisfies the NAI property. This can be seen, for instance, by taking functions that go to infinity in a small interval but are near the center in the rest of the domain.

The NAI notion is not very informative if $\| X\|$ is bounded with $ \mathbb{P}-$probability one. Therefore, we generalize it to the concept of `null at the boundary' (NAB) which is defined in terms of quantiles rather than norms of the functional observations. The formal definition is given in Appendix where it is also shown that ED satisfies NAB property. Although BD may satisfy convexity and NAB properties, it may do so trivially  due to the degeneracy problem noted earlier. ID and MBD do not satisfy NAB, since they do not satisfy the weaker NAI property.  \vspace{-2ex}

\subsection{  Convergence of Sample ED }

\cite{Fraiman01} showed that, under suitable regularity conditions,  the finite-sample versions of ID converge to the population quantity. The following proposition establishes the analogous consistency result for ED under suitable regularity conditions. The conditions and proof are given in the Appendix.

\begin{prop} \label{converge-thm}
Let $\mathbb{P}$ be a stochastic process satisfying the regularity conditions \ref{eq-cond}  - \ref{cont-cond} in the Appendix. Let $\mathbb{P}_n$ be the empirical distribution based on $n$ samples from $\mathbb{P}$. Then, \vspace{-2ex}
\[ \lim_{n \rightarrow \infty} \sup_{f \in C[0,1]} |ED(f, \mathbb{P}_n) - ED(f, \mathbb{P})| \rightarrow 0,\]
\end{prop}

\subsection{Non-Degeneracy of ED}

\cite{Chakraborty12} showed that several existing notions of functional depth suffer from the following degeneracy problem. For a general class of continuous time Gaussian processes, with probability one, the depth of every function is zero. This is true for BD and the extensions of projection depth and half-region depth to functional data in the literature. ID, MBD, and SD do not suffer from these problems. Proposition \ref{nondeg-thm} shows that extremal depth is non-degenerate for a general class of stochastic processes.

Consider $X = \{ h(t, Y_t)\}, t \in [0,1] $, where: i) $Y_t$ is a mean zero Gaussian process having  continuous sample paths, bounded variance function $0 < \sigma^2(t)  := E(Y^2(t))  <\infty$, and $\sup \{Y_t/\sigma(t), {t \in [0,1]} \}$ has a continuous distribution, ii) the function $h:[0, 1]\times \mathbb{R}$ is continuous, and iii) $h(t, .)$ is strictly increasing with $h(t, s) \rightarrow \infty$ as $s \rightarrow \infty$ for each $t \in [0, 1].$ Let $X \sim \mathbb{P}$, and define the \emph{range of ED} for $X$ as $R:= \{\alpha \in [0,1]: ED(f, \mathbb{P}) = \alpha, \text{ for some }  f \in C[0, ~1]  \}$.
Then:

\begin{prop} \label{nondeg-thm}
The range of ED for $X$ is $(0,1]$.
\end{prop}
The result is proved in the Appendix.

\section{ Central Regions Based on ED} \label{central-sec}
This section deals with construction of ED-based central regions, their theoretical properties and comparison with central regions based on other depth notions.  \vspace{-2ex}

\subsection{Definition and Properties}
Consider a function space $S$ of interest (such as $C[0,1]$ or a sample of $n$ functional observations), and let  $\mathbb{P}$ be the associated distribution of interest. 
Let $(1- \alpha)$ be the desired coverage level, {  where coverage of a region $C$ is given by $\mathbb{P}[f:  \inf \limits_{g \in C} g(t) \leq f(t) \leq \sup \limits_{g \in C} g(t), \forall t \in [0,1]$.]} Define the lower and upper $\alpha-$envelope functions as  \vspace{-2ex}
\begin{equation} \label{central-eq}
\begin{array}{clll}
f_L(t) := \inf \{f(t): f \in S,~ ED(f, \mathbb{P}) > \alpha \},\\   f_U(t) := \sup \{f(t): f \in S,~ ED(f, \mathbb{P}) > \alpha \},  \vspace{-2ex}
\end{array}
\end{equation}
respectively. Then, the $(1 - \alpha)$ ED central region is given by  \vspace{-2ex}
\begin{equation} \label{envelopes}
C_{1 - \alpha} =  \{f \in S:  f_L(t) \leq f(t) \leq  f_U(t), \forall t \in [0,1]\}.  \vspace{-3ex}
\end{equation}
When $S$ is a finite set of functions and $\mathbb{P}$ is the empirical distribution, then $C_{1 - \alpha}$ is just the convex hull formed by all the sample functions having depth larger than $\alpha$. { When $S$ is $C[0,1]$, and the marginal distribution of $\mathbb{P}$ at $t$ has zero mass to the right of $f_L(t)$ or to the left of $f_U(t),$ we take $f_L (t)$ and $f_U (t)$ to be the largest and smallest possible values (which retain the marginal probability of the interval $[f_L(t), f_U(t)]$), respectively. }

The following proposition shows that the central region of level $\alpha$ contains at least the desired amount of coverage $(1- \alpha)$. Further, when the boundary of the central region does not have any mass, the actual coverage equals the desired coverage exactly. This property is not shared by ID or MBD, and they often tend to have over-coverage problem. The proof is provided in the Appendix.

Fix $\alpha$ in the range of ED. Define the boundary set of $C_{1 - \alpha}$ as $\partial C_{1 - \alpha} = \{f \in C_{1 - \alpha}:  f(t) = f_L(t) \text{ or } f_U(t) \text{ for some } t \in [0,1]\}$. Then:

\begin{prop} \label{cover-lem}
We have\vspace{-2ex}
\begin{equation}
1 - \alpha \leq \mathbb{P} \left[ f \in C_{1 - \alpha} \right] \leq  (1 - \alpha) + \mathbb{P} \left[f \in \partial C_{1- \alpha} \right]. \vspace{-2ex}
\end{equation}
In particular,  if $\mathbb{P} \left[f \in \partial C_{1 - \alpha} \right] = 0$, we have $\mathbb{P} \left[   C_{1 - \alpha} \right] =  1 - \alpha$.  \vspace{-1ex}
 \end{prop}
As we shall see in Section \ref{more-apps}, this property is very useful in achieving desired coverage in simultaneous inference problems. When $S$ is the set of $n$ sample functions, the boundary set $\partial C_{1 - \alpha}$ is the same as the set of functions in $C_{1 - \alpha}$ that equal $f_L$ or $f_U$ (defined in Equation \eqref{envelopes}) for a part of the domain.  The probability $\mathbb{P} \left[f \in \partial C_{1 - \alpha} \right]$ may not be  exactly zero in finite samples if there are one or more functions $f_i(t)$ which coincide with the upper or lower envelopes of the central region over an interval. However, in most situations of interest, this probability goes to zero as $n \rightarrow \infty$.

ED central regions have another interesting and attractive property: there is a close relationship between the ED (simultaneous) regions and the usual pointwise central regions. Specifically, for a fixed $\gamma \in (0, 1)$, let $Q_{1 - \gamma}$ be the  $(1 - \gamma)$- pointwise central region given by  \vspace{-2ex}
\begin{equation}
Q_{1 - \gamma}  =  \{f \in S: q_{\gamma/2} (t) \leq  f(t) \leq q_{1 - \gamma/2} (t), \forall t \in [0,1].  \vspace{-2ex}
\end{equation}
Here $q_{\eta} (t)$ is the $\eta-$th quantile of the univariate distribution of $\mathbb{P}$ at $t$. Then, it is shown below that for every $\gamma \in [0, 1]$, $Q_{1 - \gamma}$ corresponds to an ED central region for some $\alpha$. Thus, every pointwise central region is an ED central region. 
\begin{prop} \label{prop-width-thm}
Let $\mathbbm{P}$ be the stochastic process of interest. For any $\gamma  \in [0,1]$,  there exists an ED central region $C_{1 - \alpha}$ for some $\alpha$ such that
$\mathbbm{P} \left[  Q_{1- \gamma} ~ \Delta ~ C_{1- \alpha} \right] = 0 $, where $\Delta$ denotes set difference. That is, up to a set of $\mathbbm{P}-$measure zero, the two sets are the same.  \vspace{-1ex}
\end{prop}
 Note that while $Q_{1-\gamma}$ corresponds to an ED central region for each $\gamma$, the converse may not be true in general. However, there is indeed a one-to-one correspondence for most continuous stochastic processes. For example, let $X = \{ h(t, Y_t)\}, t \in [0, 1],$ where $Y_t$ and $h( \cdot, \cdot)$ satisfy the conditions in Proposition \ref{nondeg-thm}.

\begin{cor} \label{loc-scale}
For every ED central region $C_{1 - \alpha}$ of $X$, there exists $\gamma \in [0,1]$ such that $\mathbbm{P} \left[ C_{1 - \alpha}~ \Delta ~ Q_{1 - \gamma} \right] = 0 $. In particular, all the ED central regions for  $Y := \{Y_t, t \in [0,1] \}$ take the form $\{ f:  -w \sigma(t) \leq f(t) \leq w\sigma (t) ,  \forall t \in [0,1] \}$, for some $w >0$.
\end{cor}
The last statement of Corollary \ref{loc-scale} implies that ED central regions for the Gaussian process $Y$ have width proportional to the standard deviation, which are perhaps the most natural central regions. {  Although ED and resultant central regions do not take explicitly into account the dependence structure of the underlying processes, the properties of the ED central region will still depend on the covariance structure.  That is, even though two different Gaussian processes may have the same point-wise variance, their $(1 - \alpha)$ ED central regions would be different depending on their covariance structure. This is because the covariance structure would determine how much width $w$ is needed to have $(1 - \alpha)$ coverage.}

\subsection{  Comparison of Central Regions} \label{realdata-ex}

\begin{figure}[!ht]
\caption{Central regions of Orthosis data set: 90 \% and 50 \% central regions in the upper and lowe panels, respectively. \label{orthosis-central}}
\begin{center}
\includegraphics[width=6.1in,height=2.1in]{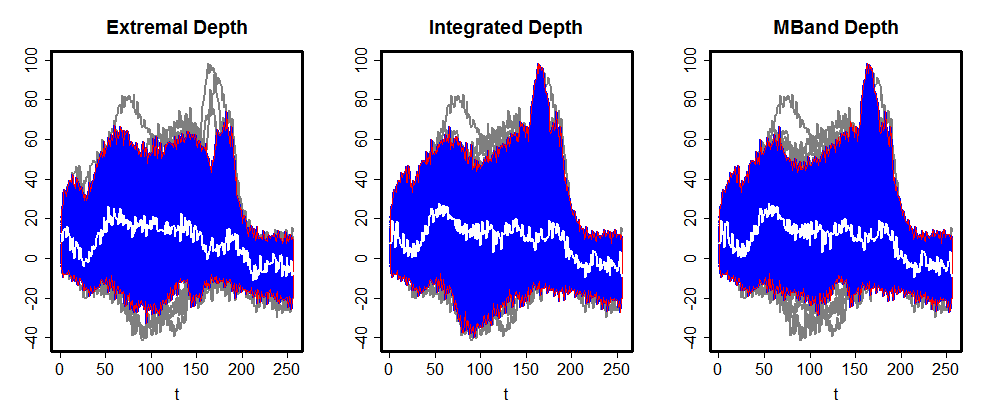}
\includegraphics[width=6.1in,height=2.1in]{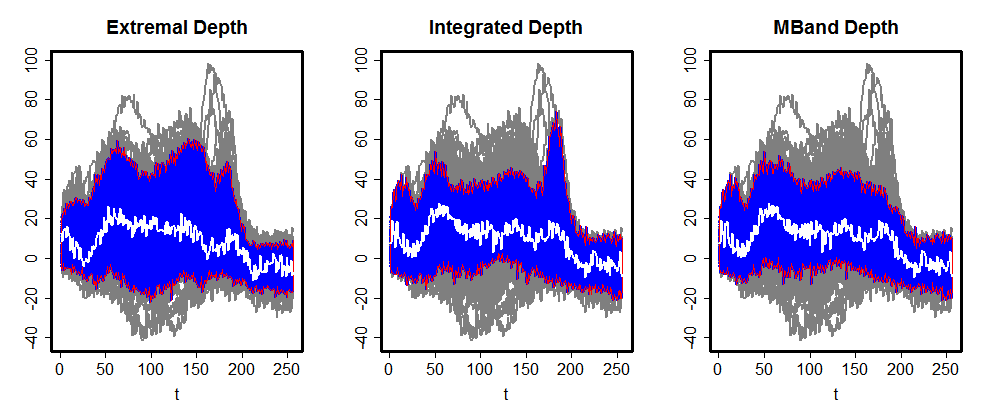}  \vspace{-2ex}
\end{center}
\end{figure}

We use the orthosis dataset considered earlier to compare the central regions formed by ED with those from other functional depths. Figure \ref{orthosis-central} compares the $90 \%$ (upper panel) and $50 \%$ (lower panel) regions formed by ED, ID and MBD. The ID and MBD central regions are defined in a similar way as the ED central regions: the convex hull formed by the deepest $(1 - \alpha) \times 100 \%$ of the sample functions.

In the upper panel, both ID and MBD regions include the peak (at the top) at around the value of 180 on x-axis while ED does not. The ID region in the lower panel ($50\%$) also includes some of this peak. Of course, one does not know the ``right'' answer in this case. However, the connection with pointwise intervals would suggest that behavior of the ED regions is more reasonable.

\begin{figure}[!ht]
\caption{Width of the 90 \% and 50 \% central regions using different approaches: The blue dots are the widths versus standard deviation and the solid black line is the least squares line. It can be seen that the ED  has width mostly proportional to the standard deviation while having relatively smaller or comparable width. \label{orthosis-width}}
\centering
\includegraphics[width=6.3in,height=2in]{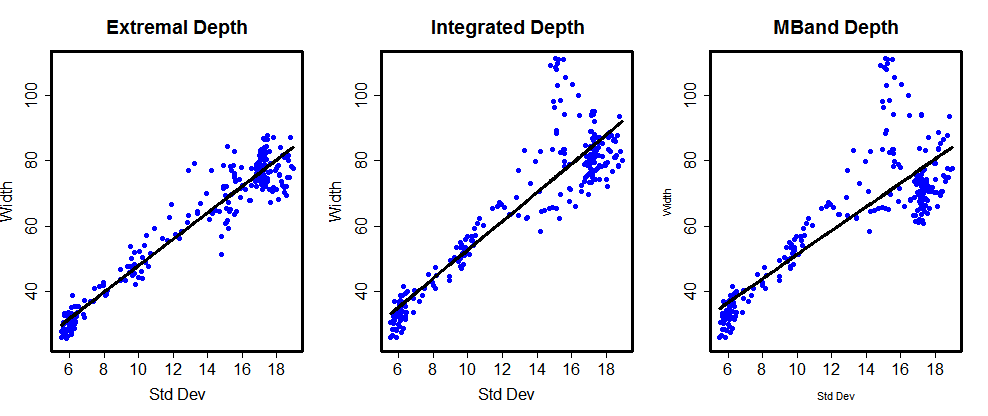}
\includegraphics[width=6.3in,height=2in]{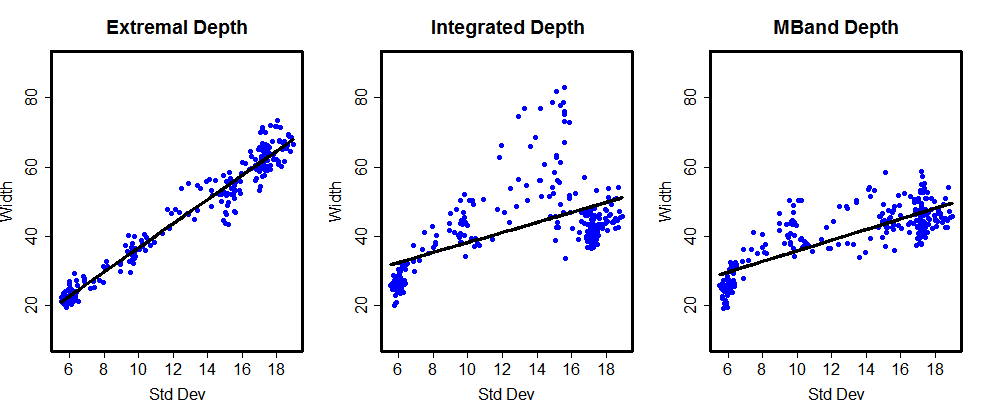} \vspace{-2ex}
\end{figure}

Figure \ref{orthosis-width} is a plot of the widths of the ED, ID, and MBD central regions against the pointwise standard deviations of the data.  We see that the ED central regions scale (approximately) proportionally to the pointwise standard deviations. This is not the case for the regions based on ID or MBD.

\section{Functional Boxplots and Outlier Detection} \label{fbplot-sec}
 \vspace{-2ex}
\subsection{Boxplots} \label{fbp-sim}

Central regions can be readily used to construct functional boxplots that provide a summary of the  data. \cite{Sun11} used MBD to develop functional boxplots that are analogous to classical boxplots for univariate data. The plot includes middle $50\%$ central region (the `box') and an envelope obtained by inflating the central 50\% central region by 1.5 times its pointwise range, the boundaries of which are referred to as `whiskers'. Functions outside this envelope are considered potential candidates for outliers.

We use a simulation study to compare the performance of ED-based functional boxplots to those based on MBD (\cite{Sun11})  and ID. The models considered below in our analysis are the same as those in \cite{Sun11}. \vspace{-2ex}\\

\noindent {\bf Model 1:} \emph{Baseline}: $X_i (t) = 4t + e_i (t ), 1 \leq i \leq n$, where $e_i (t)$ is a Gaussian process with mean zero and covariance function $\gamma  (s, t) = \exp \{-|t -s|\}$. This is the baseline model for the subsequent models. \vspace{-2ex}\\

\noindent Models $2 - 5$ include outliers. Here $\{ c_i, i =1 \leq i \leq n \}$ are indicator functions of outliers and are $i.i.d$ Bernoulli with $ p =0.1.$ That is, on average $10 \%$ of the observations are outliers. $\{ \sigma_i, i =1 \leq i \leq n \}$ are variables that take on values $\pm 1$ with equal probability and indicate the direction of the outliers; $K = 6$ is the magnitude of the outlier. \vspace{-2ex} \\

\noindent {\bf Model 2:} \emph{Symmetric contamination}: $Y_i (t) = X_i (t) + c_i \sigma_i K$. \vspace{-2ex} \\

\noindent {\bf Model 3:} \emph{Partial contamination}: Let $T_i$ be randomly generated from uniform distribution on $[0, 1]$. Then, $Y_i (t) = X_i (t) + c_i \sigma_i K$, if $t \geq T_i$ , and $Y_i (t) = X_i (t)$, if $t <T_i$. \vspace{-2ex} \\

\noindent {\bf Model 4:} \emph{Contaminated by peaks:} Let $T_i$ be randomly generated from uniform distribution on $[0, 1-\ell]$. Then, $Y_i (t) = X_i (t) + c_i \sigma_i K$, if $T_i \leq t \leq T_i + \ell$, and $Y_i (t) = X_i (t)$ otherwise. In the simulation, we fixed $\ell = 0.08$. \vspace{-2ex}\\

\noindent {\bf Model 5:} \emph{Shape contamination with different parameters in the covariance function}: $Y_i(t) = 4 t + \tilde{e}_i(t)$, where $\tilde{e}_i$ is a mean zero Gaussian process with covariance  $\gamma (s, t) = k \exp \{-|t - s|^{\mu}\}$, with $k =8, \mu = 0.1$. \vspace{-2ex}\\

\begin{figure}[h]
\begin{center}
\caption{Functional boxplots: The top and bottom panels correspond to data from Models 3 and 4, respectively. In each plot, the region in blue is the central $50 \%$ region and the lines in red are the whiskers.  \label{model3}}

\includegraphics[width=6.2in,height=2.3in]{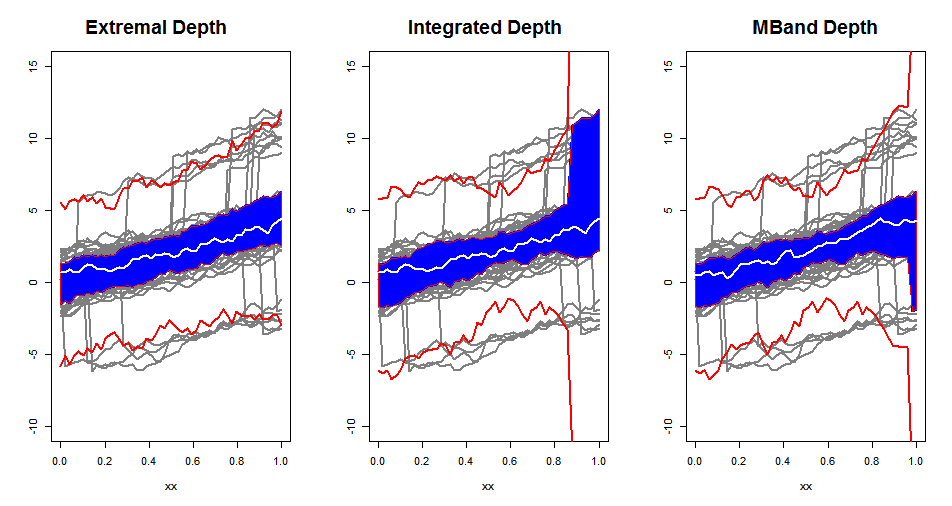}
\includegraphics[width=6.2in,height=2.3in]{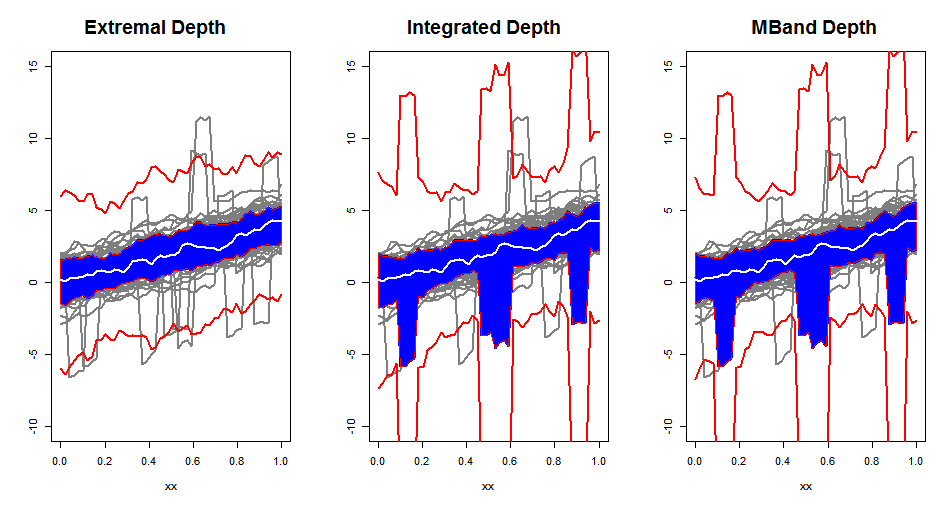}  \vspace{-2ex}
\end{center}
\end{figure}

For the simulation, we generated $n = 100$ functional observations from the above models and evaluated them on a grid of size $50$. Only a summary of the results is given here. For the baseline model with no outliers, all of the depths lead to `well-behaved' boxplots. With outliers, ID and MBD-based boxplots exhibited undesirable features, and this was most evident for Models 3 and 4. Figure \ref{model3} shows a sample dataset. For Model 3 (upper panel), the middle $50\%$ of the central region is affected by the $10\%$ contamination. The problem is less so for MBD but it is still evident. The issue is more serious for Model 4 where the performances of both ID and MBD are badly affected. As noted, part of the reason is that both ID and MBD rely on some type of averaging. The ED plots, which rely on the extremal property, are unaffected by the outliers in these examples.  \vspace{-2ex}

\subsection{Outlier Detection} \label{outlier}

This section provides a formal comparison of the performance of boxplots as outlier-detection tools. We use the same measures in \cite{Sun11} for comparison:\\
i) $p_c$: percentage of correctly identified outliers, and \\
ii) $p_f$: percentage of incorrectly detected outliers (equals the number of incorrectly identified outliers divided by total number of non-outlying functions). The standard errors of the percentages are given in parenthesis.

Table \ref{roc-like-table} shows the results based on 100 data sets simulated using the Models 1-5 described above. We see that $p_f-$values of ED are much lower across all models. The values of $p_c$ are generally similar for the different depth notions except for model 4, where ED outperforms by a clear margin. This is not surprising as model 4 is contaminated by peaks; ID and MBD fail to find the outliers due to their ``averaging" property as was evident in Figure \ref{model3}.

These results suggest that when the outlying functions are consistently outlying in the whole domain, all three notions -- ED, ID and MBD -- perform well. However, when there are functions that are outlying in a subset of the domain as in Models 3 and 4, ED performs better while ID and MBD can do poorly.

\begin{table}[ht]
\caption{Outlier detection using Functional Box-Plots: $p_c$ is the percentage of correctly identified outliers; $p_f$ is the proportion of incorrectly identified outliers. Numbers in brackets indicate their standard errors.  \label{roc-like-table}}
\centering
\begin{tabular}{ccccccc}
  \toprule
 & &ED&ID&MBD& \\
\midrule
    Model 1   & $p_f$ & {\bf 0.03 } (0.17) &0.06 (0.25)&0.07 (0.27) \\
  \midrule
     \multirow{2}{*}{Model 2} & $p_c$ &98.52(4.42) &98.89(3.49)& {\bf 99.15}(3.03)  \\
      & $p_f$ & {\bf 0.01 } (0.10) &0.03 (0.20)&0.04 (0.21) \\
  \midrule
  \multirow{2}{*}{Model 3} & $p_c$ &{\bf 86.43}(13.64) &77.24
(16.72)&83.17(13.77)  \\
      & $p_f$ &{\bf 0.01} (0.12) &0.03 (0.18)&0.03 (0.21) \\
  \midrule
   \multirow{2}{*}{Model 4} & $p_c$ &{\bf 84.42 }(13.29) &41.06
(18.90)&45.94(18.99)  \\
      & $p_f$ &{\bf 0.01} (0.17) &0.04 (0.21)&0.04 (0.22) \\
  \midrule
   \multirow{2}{*}{Model 5} & $p_c$ &75.74(16.15) &74.97
(16.91)& {\bf 78.17} (15.79)  \\
      & $p_f$ & {\bf 0.01 } (0.11) &0.03 (0.19)&0.04 (0.24) \\
   \bottomrule

\end{tabular}
\end{table}

The above discussion indicates that the corresponding estimators, such as functional trimmed means, based on ED will be more resistant to outliers. Specifically, let $m(\alpha)$ is the trimmed mean based on the sample functions in $(1-\alpha)$ ED central region. Then, the simulation results suggest that $m(\alpha)$ may remain bounded even as the outliers increase in magnitude while the corresponding trimmed means for ID and MBD can be unbounded. This result can be established formally and we plan to pursue this in the future. \vspace{-2ex}

\section{Simultaneous Inference}\label{more-apps}

 In problems involving functional inference, such as regression and density estimation, it is often difficult to obtain exact simultaneous confidence bands. In such cases, one can combine resampling methods, such as the bootstrap \citep{Efron79}, with central regions using functional depth  to obtain simultaneous confidence regions. Under the asymptotic validity of the resampling technique, we can get approximate simultaneous confidence regions of desired coverage. This section demonstrates the application for the case of polynomial regression and compares it with other methods.  \vspace{-2ex}

\subsection{Polynomial and Other Parametric Regression}\label{poly-sec}

Consider the polynomial regression problem $Y(x_i) = \mu(x_i)+ \epsilon_i$ with $\mu(x_i) = \theta_0 +  \theta_1 x_i  + \cdots+ \theta_q x_i^q$.
The covariates $x_i$'s are fixed and the error terms $\epsilon_i$'s are $i.i.d$ with the standard regression assumptions. The goal is to get a simultaneous confidence region for $\mu(x)$ for all $x$.

It is known that there is no `exact' method for this general problem. Scheffe's method  \citep{Scheffe59} leads to conservative regions since a polynomial of a variable $x$ of degree $q$ does not span the full $(q+1)-$dimensional Euclidean space.
The level of conservatism gets higher as the degree $q$ increases. Exact methods have been developed in special cases. \cite{Piegorsch86} considered quadratic regression and provided confidence bands sharper than Scheffe's bands. \cite{Liu14} proposed exact bands for quadratic and cubic polynomial regressions. \cite{Wynn84} developed exact bands when the errors are normally distributed using special properties of normality. We describe here a general re-sampling based approach using ED central regions.

Let $\theta = (\theta_0,\theta_1, ... \theta_q)$ denote the vector of parameters, $\hat{\theta}$ denote the usual least-squares estimator, and $\hat{\mu}(x)$ be the corresponding predictor. Consider the residuals $r_i = Y(x_i) - \hat{\mu}(x_i)$ and $\hat{s}$, the residual standard error. Generate $B$ bootstrap samples from the residuals to obtain bootstrap estimates $\hat{\theta}^*_1, \hat{\theta}^*_2, \cdots, \hat{\theta}^*_B$ of $\theta$, and  $\hat{s}^*_1, \hat{s}^*_2, \cdots, \hat{s}^*_B$, of $\sigma$. Define an estimate of the polynomial mean function  $\hat{\mu}(x|\hat{\theta}^*)$ in the obvious manner and the normalized (centered and scaled) version of this function as \vspace{-2ex}
\begin{equation} \label{bootstrap-fn}
m^*_j (x)= \frac{\hat{\mu}(x|\hat{\theta}^*_j) - \hat{\mu}(x|\hat{\theta})}{\hat{s}^*_j},
\end{equation}
for $j = 1, 2, \cdots B$.
These are pivotal quantities: their distribution is free of $\theta$ and $\sigma$. The set of normalized bootstrapped functions $S^*: = \{m^*_1, m^*_2, \cdots, m^*_B \}$ can now be treated as our functional data, and they can be used to construct the ED central region. Specifically, let $f_L^*(x)$ and $f_U^*(x)$ be the lower and upper envelopes of this region. Then, the $(1 - \alpha)-$level simultaneous confidence band for $\mu(x)$ is given by \vspace{-2ex}
\begin{equation}
C_n^{\alpha}  = \{\mu(x): \hat{\mu}(x) + \hat{s} f_L^*(x) \leq \mu(x) \leq \hat{\mu}(x)+ \hat{s} f_U^*(x), \forall x \}. \vspace{-2ex}
\end{equation}
Based on the results in Section \ref{central-sec}, and the bootstrap validity for parametric regression models \citep{Freedman81}, we get $P[\mu(x) \in C_n^{\alpha}  ~\forall ~x] \rightarrow (1 - \alpha)$ as $n \rightarrow \infty$.

\begin{figure}[h]
\begin{center}
\caption{Simultaneous confidence bands: The  figure on the left plots all the bootstrapped functions along with 90 \% ED central region and the  plot on the right gives confidence bands from the three different methods \label{poly-fig}}
\includegraphics[width=5.4in,height=2.4in]{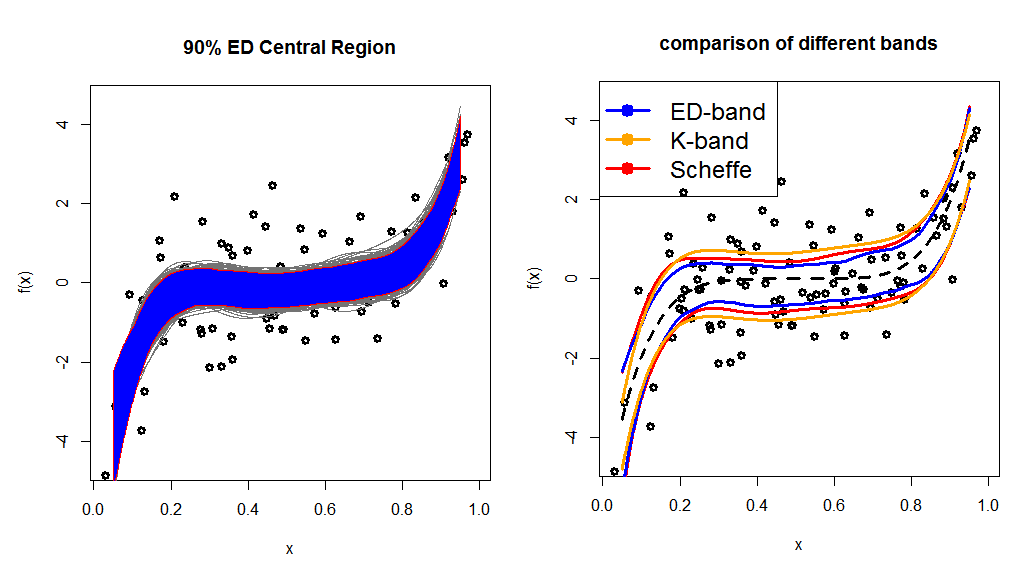}  \vspace{-4ex}
\end{center}
\end{figure}

We use a limited simulation study to examine the finite sample performance of this band and compare it with bands based on Scheffe's method and a Kolmogorov-like sup-norm statistic. The sup-norm statistic is $K_j^*  = \sup_{x} (|\hat{\mu}_j^*(x) - \hat{\mu}(x)|)/\hat{s}_j^*$. The Scheffe's band is obtained in the usual manner assuming normality. The simulation was done for a degree five polynomial $\mu(x) =192 (x - 0.5)^5$; the coefficient 192 was chosen so that the absolute mean function integrates to one. This is the dashed function in the right panel of Figure \ref{poly-fig}. We simulated $n=100$ observations with $i.i.d.$ normal error terms having standard deviation 5; the covariate $x$ was randomly generated from $U[0, 1]$. We used $B = 2000$ bootsrap samples for obtain ED confidence bands.

Figure \ref{poly-fig} shows the confidence bands and the true mean function (dashed line) for one data set. The confidence band based on ED are tighter than both Scheffe's and K- bands (the band using $K_j^*$'s).
Table \ref{poly-tab} gives the numerical results from the simulation study. The first row is the coverage probability and the next five rows show the power values for five different alternative polynomials.  The first two alternatives are given by $P_k = C_k ~ sign(x-0.5)~ (x - 0.5)^k $ for degrees $k = 4, 6$, and $C_k$ is a constant such that $|P_k|$ integrates to one. The next three alternatives are additive shifts from the original mean function $P_5.$

As expected, the Scheffe-band is very conservative (actual coverage is $99 \%$ while the nominal coverage is only $90 \%$). The ED-band has coverage very close to $90 \%$ as desired. The coverage proportion of the K-band is close to the nominal. However, the band is wide in the middle and narrow in the tails. This leads to lower power than the ED-band for a large class of alternatives which have shift in the middle of the domain. This can be seen in last three rows of Table \ref{poly-tab}, where K-band has substantially lower power for the three shift alternatives. Power of ED-bands for the $P_4$, $P_6$ alternatives, which mostly differ from K-band at the tails, also remains competitive. 
\begin{table}[h]
\caption{Level (row 1) and Power (rows 2 - 6) for 90 \% simultaneous confidence bands using different methods 
\label{poly-tab}} 
\centering 
\begin{tabular}{cccccc} 
\hline\hline 
  &  Scheffe   & K-band & ED\\ 
\hline
Level ($P_5$) &0.01&0.10&0.10\\
$P_4$ &0.02&0.14&0.16\\
$P_6$ &0.03&0.17&0.19\\
 $0.2+ P_5 $ &0.08&0.21&0.32\\
$0.2+ 0.2x + P_5$& 0.31&0.32&0.66\\
$0.2~ sign(x-0.5) + P_5$& 0.09&0.22&0.38\\
\hline
\end{tabular}
\end{table}

This application to polynomial regression can be readily extended to more general models of the form
 $Y_i = \theta_0 +  \phi_1({\bf x}_i) \theta_1  + \cdots + \phi_q({\bf x}_i) \theta_q + \epsilon_i,$ where $\phi_1, \cdots, \phi_q$ are splines or other known basis functions. The covariates can also be multidimensional in this framework. 

\subsection{Other Functional Inference Problems}

The resampling approach described in the last section can be readily extended to other problems. These include the goodness-of-fit testing problem where one wants to determine if the generative model belongs to a certain parametric family of distributions. One can combine the bootstrapping technique (parametric or nonparametric) with ED central regions to construct acceptance or confidence regions. (See \cite{Cuevas06}, \cite{Yeh96} and \cite{Yeh97} for some related discussion.) While this is a classical problem, our initial studies suggest that the ED-based approach has some advantages over methods based on weighted Kolmogorov statistics.

The approach can also be used to construct confidence bands in nonparametric functional estimation problems, such as regression or density estimation. However, the justification of the ED-based regions in general function estimation problems will depend on the limiting distributions of the functional estimators and the asymptotic validity of the bootstrap. Simulation results in finite samples suggest that the convergence of the actual level to the nominal one is slow in fully nonparametric inference problems. A more extensive study is needed to understand the behavior, both theoretically and empirically.  \vspace{-2ex}

\section{Concluding Remarks}

An important class of problems deals with the case where the underlying functions of interest are observed with error. In other words, instead of observing random functions $X_i(t)$ from a generative model of interest, we observe $Y_i(t) = X_i(t) + \epsilon_i(t), ~~~ i = 1, \cdots, n$. A natural approach is to use some type of smoother to `recover' $X_i(t)$ and then use the techniques discussed so far. If there is some information of the error structure in $\epsilon(t)$, this can be used to guide the smoothing algorithm or the `reconstruction' methods for $X_i(t)$.

In summary, we have developed a new notion of functional depth, studied its properties, and demonstrated its usefulness through several applications. While no single notion of functional depth will do uniformly better than others, we hope that the results here suggest that the extremal-depth concept has many attractive properties and is a useful tool for exploratory analysis of functional data. It can also be used in other applications, such as the construction of simultaneous confidence bands.  \vspace{-2ex}

\bibliographystyle{apa}
\bibliography{bibfile}
  \vspace{-3ex}
\section*{Appendix \vspace{-1ex}}
{ {More general definition of extremal depth:\\}
The depth ordering defined in the paper can be generalized as follows.  Consider a pair of functions $g, h \in C[0,1]$, and define
\begin{equation}\label{dijplus}
r^* =   \inf \{r \in [0,1]: \Phi_{g}(r)  \neq \Phi_{h}(r) \}.
\end{equation}
Then, we say $h \prec g$ if for any sequence $r_m \downarrow r^*$, we have $\Phi_{h}(r_m)  > \Phi_g(r_m)$ for all $m \geq M,$ for some large enough constant $M$.  Similarly, we say  $h \succ g$ if for any sequence $r_m \downarrow r^*$, we have $\Phi_{h}(r_m)  < \Phi_g(r_m)$ for all $m \geq M.$ Otherwise, if neither of these cases holds, we say $h \sim g$. This generalizes the definition that assumes the existence of a neighborhood in which either $\Phi_{h}(r_m)  > \Phi_g(r_m)$ or $\Phi_{h}(r_m)  < \Phi_g(r_m)$ holds true.}
\subsubsection*{Condition and proof for Proposition \ref{prop1}:  \vspace{-1ex}}
\begin{condition}\label{eq-cond} Assume that
(a) $\mathbbm{P}[d_f = 0] = 0$, and (b) $\mathbbm{P}[d_f = d_g, f \neq g] = 0,$
where $f, g$ are independent random functions from $ \mathbbm{P}$ and $d_f : = \inf \limits  \{r \in [0,1]: \Phi_f(r ) >0 \}$.
\end{condition}

\noindent {\bf Proof of Proposition \ref{prop1}:} We shall show that, if $ED(f_1, \mathbb{P}), ED(f_2, \mathbb{P}) \geq \alpha$, and $f_1(t) \leq f(t)  \leq f_2(t)$ $\forall t \in [0,1]$, then $ED(f, \mathbb{P}) \geq \alpha.$ Note that $\forall t$, $D_{f} (t, \mathbb{P}) \geq \min (D_{f_1} (t, \mathbb{P}), D_{f_2} (t, \mathbb{P}))$, and hence $d_f \geq \min(d_{f_1}, d_{f_2})$. Therefore either $f \succeq f_1$ or $f \succeq f_2$ w.p.1 and $f \in C_{\alpha}$ due to Condition \ref{eq-cond} (b). \qed

Condition \ref{eq-cond} is a mild condition on $\mathbbm{P}$. For instance, this holds if $ \mathbb{P}$ is the distribution of $X$ in Proposition \ref{nondeg-thm}.  \vspace{-2ex}
\subsubsection*{NAB property: \vspace{-1ex}}
Denote the pointwise quantile functions of $\mathbb{P}$ as $q_{\alpha}$, i.e., for each $t$, $\mathbb{P}[X(t) < q_{\alpha}(t)] \leq \alpha$, and $\mathbb{P}[X(t) \leq q_{\alpha}(t)] \geq \alpha$ (for uniqueness we take the smallest one).
Let for $\alpha_n \downarrow 0$, $f_n(t) \leq q_{\alpha_n}(t), \forall t \in U$, and for $\beta_n \uparrow 1$, $g_n(t) \geq q_{\beta_n}(t), \forall t \in U$, where $U$ is some open interval in $[0,1]$. Then we say the depth notion $D$ to have NAB property if $D(f_n, X) \rightarrow 0$ and $D(g_n, X) \rightarrow 0$.

We now show that ED satisfies NAB under Condition  \ref{eq-cond} (a). Since $f_n(t) \leq q_{\alpha_n}(t), \forall t \in U$, we have $\forall n \geq N$, \vspace{-2ex}
\begin{equation*}
\begin{array}{lllllllll}
\mathbb{P}[f_{n+1} \succeq X] & \leq & 1 - \mathbb{P}[ q_{\alpha_n} < X < q_{1 - \alpha_n}]\\
& =&  1 - \mathbb{P}[ \cup_{k \leq n} \{ q_{\alpha_k} < X < q_{1 - \alpha_k} \}]. \vspace{-2ex}
\end{array}
\end{equation*}
Therefore,
$\lim \sup \mathbb{P}[f_{n+1} \succeq X] \leq 1 - \mathbb{P}[\Omega:= \cup_{1 < k < \infty} \{q_{\alpha_k} < X < q_{1 - \alpha_k} \}] = 0$, as the set $\Omega$ has probability one due to Condition \ref{eq-cond} (a). Therefore, $D(f_n, X) \rightarrow 0$ and similarly $D(g_n, X) \rightarrow 0$. \vspace{-2ex}

\subsubsection*{Conditions and proof of Proposition \ref{converge-thm}: \vspace{-1ex}}
\begin{condition} \label{cross-cond}
Let $C_n$ be the total number of functional crossings by any pair of functions, where $n$ is the number of sample functions. We assume that $C_n = \exp\{ o_{P}(n) \}$. \vspace{-2ex}
\end{condition}
\begin{condition} \label{cont-cond}
Let $\mathbb{P}$ be a stochastic process on $C[0,1]$ whose univariate CDF at $t \in [0,1]$ is denoted by $F_t$.  Define $R(\delta, u) = \sup \limits_{|t-s| < \delta} |F_t(u) - F_s(u)|$. Then we assume that for any $u_0$, there is a neighborhood $B(u_0, \epsilon)$ such that $R(\delta_n,u) \rightarrow 0$  uniformly in $u \in B(u_0, \epsilon)$ as $\delta_n \rightarrow 0$.  Further, we assume $\mathbb{P}$ to have Glivenko-Cantelli (GC) property uniformly over convex sets. \vspace{-2ex} \end{condition}

Condition \ref{cross-cond} assumes the number of crossings is at most exponential in sample size, and is related to the smoothness of the process. Condition \ref{cont-cond} assumes that the CDF's of neighboring points in the domain are close. The GC property of $\mathbb{P}$ requires that the empirical distributions corresponding to $\mathbb{P}$ converge uniformly over convex sets. GC property for convex sets  holds under general conditions for finite dimensional distributions \citep{Eddy77}. \cite{Chakraborty14} provides a GC type result for spatial distributions of infinite-dimensional spaces. For our result, we assume this as a technical condition.

Let $f \succeq_n g$ and $f \succeq g$ denote that $f$ is deeper than or equal to $g$ using ED w.r.t. the empirical distribution $\mathbb{P}_n$ and the true distribution $\mathbb{P}$, respectively. Then, \vspace{-2ex}
\begin{equation} \label{split-eq}
\begin{array}{llllllll}
 \sup \limits_{f \in C[0,1]}  |ED(f, \mathbb{P}_n) - ED(f, \mathbb{P})|  & = & \sup \limits_{f \in  C[0,1]}  |\mathbb{P}_n \left[ f \succeq_n X_n \right] - \mathbb{P} \left[ f \succeq X \right]|\\
& = & \sup \limits_{f \in C[0,1]}  |\mathbb{P}_n \left[ f \succeq_n X_n \right] - \mathbb{P} \left[ f \succeq_n X \right]| \\
& & \qquad +  \sup \limits_{f \in C[0,1]}  |\mathbb{P} \left[ f \succeq_n X \right] - \mathbb{P} \left[ f \succeq X \right]|, \vspace{-2ex}
\end{array}
\end{equation}
where $X_n \sim \mathbb{P}_n$ and $X \sim \mathbb{P}$.

The first term in RHS of \eqref{split-eq} can be shown to go to zero because of the Glivenko-Cantelli (GC) property assumed by Condition \ref{cont-cond}.  That is because the sets $\{ f \succeq_n X \}$ are convex and the GC type result holds over all convex subsets. We then only need to show that $\sup_{f  }  |\mathbb{P} \left[ f \succeq_n X \right] - \mathbb{P} \left[ f \succeq X \right]|\rightarrow 0$.

We will now show that the second term in RHS of \eqref{split-eq} goes to zero. Define $d_f : = \inf \limits_{y \in [0,1]} \{ \Phi_f(y, \mathbb{P}) >0 \}$, and $d_f^n: = \inf \limits_{y \in [0,1]} \{ \Phi_f(y, \mathbb{P}_n) >0\}.$ We shall first show that $\sup_{f} |d_f^n - d_f| \cprob 0$ as $n \rightarrow \infty$.

Due to the rate of Glivenko-Cantelli of empirical distributions \citep{Pollard91}, we have for any $t$, \vspace{-2ex}
\begin{equation} \label{gc-ineq}
\mathbb{P}  \left[ \sup \limits_{u} |F^n_t(u) - F_t(u)| > \epsilon \right] \leq \exp \{ -c \epsilon^2 n \}. \vspace{-2ex}
\end{equation}
Let $D = \{d_1, d_2, \cdots, \}$ be a countable dense subset of $[0,1]$. Define $T_n = \{ t_1, \cdots, t_{k_n}\}$ be the set containg all the points in $[0,1]$ where $n$ sample functions cross and along with $\{ d_1, d_2, \cdots, d_n \}$.  As $n \rightarrow \infty$ we have $T:= \cup_{n} T_n $ is the union of all the crossing points and $D$. Due to Condition \ref{cross-cond}, we have $\log |k_n| =  {o_P(n)}.$

Due to Equation \eqref{gc-ineq}, we have \vspace{-2ex}
\begin{equation}\label{unif-conv}
\mathbb{P}  \left[\sup \limits_{t \in T_n} \sup \limits_{u} |F^n_t(u) - F_t(u)| > \epsilon \right] \leq \exp \{ -c' \epsilon^2 n + \log k_n \}. \vspace{-2ex}
\end{equation}

Now, note that $d_f^n= \inf \limits_{t \in [0,1]} D_f(t, \mathbb{P} _n) =  \min \limits_{t \in T_n} D_f(t, \mathbb{P} _n),$ because the univariate depths in $T_n$ have the same range as that of the whole interval $[0,1]$. We shall first show that \vspace{-2ex}
\begin{equation} \label{dfeq}
d_f  =  \inf \limits_{t \in T} D_f(t, \mathbb{P} ) =\lim_{n} \inf \limits_{t \in T_n} D_f(t, \mathbb{P} ),  \vspace{-2ex}
\end{equation}
using the facts that $\cup_n T_n = T$, $T$ is dense and Condition \ref{cont-cond}. To see this, first note that $d_f \leq    \inf \limits_{t \in T} D_f(t, \mathbb{P} )$. For the reverse inequality, consider a $y_0$ such that $D_{y_0}(f, \mathbb{P} ) = d_f$ (this exists due to continuity of $F_t$ in $t$). Since $T$ is dense, we have a sequence $y_n \in T$ such that $y_n \rightarrow y_0$. Due to continuity of $F$ and Condition \ref{cont-cond}, we have \vspace{-2ex}
\begin{equation*}
\begin{array}{lllllll}
| D_{y_n} (f, \mathbb{P} ) - D_{y_0} (f, \mathbb{P} ) |& = &| |1 - 2 F_{y_n}(f(y_n))| - |1 - 2 F_{y_0}(f(y_0))||\\
 & \leq &2| F_{y_n}(f(y_n)) - F_{y_0}(f(y_0))|\\
 &\leq& 2 |F_{y_n}(f(y_n)) - F_{y_0}(f(y_n))| + 2 | F_{y_0}(f(y_n)) -  F_{y_0}(f(y_0))|  \rightarrow 0, \vspace{-2ex}
\end{array}
\end{equation*}
which implies \eqref{dfeq}. Now, using \eqref{unif-conv}, we have \vspace{-2ex}
\begin{equation*}
\begin{array}{lllllll} \mathbb{P} [\sup \limits_{f} |d_f^n - d_f| > \epsilon_n ] \leq \mathbb{P}  \left[\sup \limits_{t \in T_n} \sup \limits_{u} |F^n_t(u) - F_t(u)| > \epsilon_n/4 \right] \rightarrow 0,
\end{array}
\end{equation*}
 if  $\epsilon_n \rightarrow 0$ and $ c' \epsilon_n^2 n   - \log k_n \rightarrow \infty$. In particular, when $\epsilon_n =4 \max(\left(3 \log k_n/c'n \right)^{1/2}, 1/\sqrt{\log n})$,  $ \mathbb{P} [\sup \limits_{f} |d_f^n - d_f| > \epsilon_n]  < C  n^{-1  - \epsilon}$, for some $C, \epsilon >0$. Then using Borel-Cantelli lemma, we obtain  $\sup \limits_{f} |d_f^n - d_f| \rightarrow 0$ almost surely. Now, consider the events $A_n= \{d_f^n \geq d_g^n   \}$ and $B_m=  \{d_f < d_g - \delta_m \}$, where $\delta_m \rightarrow 0 $ as $m \rightarrow \infty$. Note that $A_n$ and $B_m$ depend on the functions $f$ and $g$. Then,
$\mathbb{P}[\cup_{f, g} A_n \cap B_m ] \leq \mathbb{P} [\sup \limits_{h} |d_h^n - d_h| > \epsilon_m] \rightarrow 0$ as $n \rightarrow \infty$. Therefore, we have
\vspace{-4ex}
\begin{equation*}
\begin{array}{lllll}
\limsup_n \sup_{f  }  |\mathbb{P} \left[ f \succeq_n X \right] - \mathbb{P} \left[ f \succeq X \right]| & \leq &
 \limsup_n \mathbb{P} \left[ \cup_{f} \{ f \succeq_n X\} \Delta \{ f \succeq X\} \right]\\
 & \leq & \limsup_n \lim_m \mathbb{P} \left[  \cup_{f, g} A_n \cap B_m \right]\\
&\leq & \lim_m \limsup_n \mathbb{P}  \left[ \cup_{f, g} A_n \cap B_m \right] = 0. \qed \vspace{-2ex}
\end{array}
\end{equation*}

\subsubsection*{Proof of Proposition \ref{nondeg-thm}:} \vspace{-2ex}
As the process $Y= \{Y_t \}, t \in [0,1]$  has continuous sample paths, the sample paths of the process $X=\{X_t \}, t \in [0,1]$ also lie in $C[0, 1]$ almost surely. Due to the monotone invariance property of ED, we only need to show that ED of $Y$ takes all the values in $[0,1]$.
Consider the sets $Q_{1-\gamma} := \{f: q_{\gamma/2}(t) \leq f(t) \leq  q_{1 - \gamma/2}(t), \forall t\},$ for $\gamma \in [0,1]$, where  $q_{\alpha}$ is the $\alpha$-th pointwise quantile of $Y$. Note that $q_{\gamma/2} \preceq f,$ for any $f \in Q_{1- \gamma}$ and $q_{\gamma/2} \succ g$, for $g \in Q_{1 - \gamma}^c$. Therefore, $ED(q_{\gamma/2},\mathbb{P} )  = \mathbb{P} [ Q_{1-\gamma}]$. By noting that $Q_{1 - \gamma} = \{f: \sup \limits_{t}| f(t)/\sigma(t)| \leq c \}$ and that $\sup \limits_{t}| f(t)/\sigma(t)|$ has a continuous distribution, $\mathbb{P} [ Q_{1 - \gamma}]$ takes all  the values in $(0,1]$. \qed \vspace{-2ex}

\subsubsection*{Proof of Proposition \ref{cover-lem}:} \vspace{-2ex}
To prove the lower bound, consider a function $g$ having ED equal to $\alpha$, { that is, $g$ is such that $ED(g, \mathbb{P}) = \alpha$. Consider the set $A:= \{ f: f \succ g, \text{ and } f \in C_{1 - \alpha}^C \}$, then $\mathbb{P}[A] = 0$. This is because, otherwise if $\mathbb{P}[A] >0$, there exists a function $f_* \in A$ such that $ED(f_*, \mathbb{P}) > \alpha$, which is a contradition as $f_* \not \in C_{1 - \alpha}$.} This implies that $\alpha = \mathbb{P}[X: g \succeq X] \geq  \mathbb{P} [ C_{1 - \alpha}^C ] $, and hence $ \mathbb{P} [ C_{1 - \alpha}] \geq (1 - \alpha) $.

To prove the upper bound, we first note that the set $\{f: g \succeq f \}$ is contained in the union of the sets $C_{1 - \alpha}^C$ and $ \partial C_{1 - \alpha}.$ This is because, for any function $h \in C_{1-\alpha} -  \partial C_{1 - \alpha},$ $d_{min}(h, \mathbb{P}) > d_{min}(g, \mathbb{P}),$ where $d_{\min}(h, \mathbb{P}) = \inf_{t \in [0, 1]} D_{h}(t, \mathbb{P})$ as in Section \ref{ed-sec}. Otherwise, we have a function $f$ with ED larger than $\alpha$ but $d_{min}(f, \mathbb{P}) <d_{min}(g, \mathbb{P}),$ which is a contradiction. Therefore, $\alpha = \mathbb{P}[X: g \succeq X ] \leq \mathbb{P}[C_{1 - \alpha}^C \cup \partial C_{1 - \alpha} ].$  This implies that $\mathbb{P}[C_{1 - \alpha} - \partial C_{1 - \alpha} ] \leq (1 - \alpha)$ and $ \mathbb{P}[C_{1 - \alpha}] \leq (1 - \alpha) + \mathbb{P} [\partial C_{1 - \alpha} ] $, and the result follows. \qed \vspace{-2ex}

\subsubsection*{Proof of Proposition \ref{prop-width-thm} \& Corollary \ref{loc-scale}:}  \vspace{-2ex}
 We shall show that the  ED central region $C^*$ formed by the functions $\{f: ED(f, \mathbbm{P}) \geq   ED(q_{\gamma/2}, \mathbbm{P})\}$ proves the proposition. Although this central region is not in the form defined by Equation \eqref{central-eq} (due to ``$\geq$'' instead of a ``$>$''), this does not make a difference when $\mathbbm{P}$ is a continuous stochastic process, and this same set can be written with a ``$>$'' when $ \mathbbm{P}$ is an empirical distribution.  We have $ f \succeq q_{\gamma/2} \sim q_{1 - \gamma/2} \succ g$, for any $f \in Q_{1 - \gamma}$, and $g \in Q_{1 - \gamma}^C$. Therefore, $Q_{1 - \gamma} \subset C^*$ and it remains to show that $\mathbbm{P}[C^* - Q_{1 - \gamma}] = 0$. However, $C^* - Q_{1 - \gamma} \subset B:= \{f \not \in Q_{1 - \gamma}: ED(f, \mathbbm{P}) =  ED(q_{\gamma/2}, \mathbbm{P})\}$. As all the functions in $B$ have the same ED, $ \mathbbm{P}[f \in B]= \mathbbm{P}[f \in B: f \sim q_{\gamma/2}] =0. $ Therefore, $\mathbbm{P}[C^* \Delta Q_{1 - \gamma}] = 0$. The corollary follows directly because ED is a decreasing function of $\sup \limits_t  |f(t)|/{\sigma(t)}$. \qed
\vspace{-2ex}
\end{document}